\def\standardrisposta{s }\def\reducedrisposta{r }
\def\mplarisposta{mpla }\def\zerorisposta{z }
\def\doublerisposta{d }\def\cartarisposta{e }\def\amsrisposta{y }
\newcount\ingrandimento \newcount\sinnota \newcount\dimnota
\newcount\unoduecol \newdimen\collhsize \newdimen\tothsize
\newdimen\fullhsize \newcount\controllorisposta \sinnota=1
\newskip\infralinea  \global\controllorisposta=0
\immediate\write16 { ********  Welcome to PANDA macros (Plain TeX,
AP, 1991) ******** }
\immediate\write16 { You'll have to answer a few questions in
lowercase.}
\message{>  Do you want it in double-page (d), reduced (r)
or standard format (s) ? }\read-1 to\risposta
\message{>  Do you want it in USA A4 (u) or EUROPEAN A4
(e) paper size ? }\read-1 to\srisposta
\message{>  Do you have AMSFonts 2.0 (math) fonts (y/n) ? }
\read-1 to\arisposta
%
%
%
%
%
\ifx\risposta\standardrisposta \ingrandimento=1200
\message {>> This will come out UNREDUCED << }
\dimnota=2 \unoduecol=1 \global\controllorisposta=1 \fi
\ifx\risposta\reducedrisposta \ingrandimento=1095 \dimnota=1
\unoduecol=1  \global\controllorisposta=1
\message {>> This will come out REDUCED << } \fi
\ifx\risposta\doublerisposta \ingrandimento=1000 \dimnota=2
\unoduecol=2   \message {>> You must print this in
LANDSCAPE orientation << } \global\controllorisposta=1 \fi
\ifx\risposta\mplarisposta \ingrandimento=1000 \dimnota=1
\message {>> Mod. Phys. Lett. A format << }
\unoduecol=1 \global\controllorisposta=1 \fi
\ifx\risposta\zerorisposta \ingrandimento=1000 \dimnota=2
\message {>> Zero Magnification format << }
\unoduecol=1 \global\controllorisposta=1 \fi
\ifnum\controllorisposta=0  \ingrandimento=1200
\message {>>> ERROR IN INPUT, I ASSUME STANDARD
UNREDUCED FORMAT <<< }  \dimnota=2 \unoduecol=1 \fi
\magnification=\ingrandimento
%
%
%
%
\newdimen\eucolumnsize \newdimen\eudoublehsize \newdimen\eudoublevsize
\newdimen\uscolumnsize \newdimen\usdoublehsize \newdimen\usdoublevsize
\newdimen\eusinglehsize \newdimen\eusinglevsize \newdimen\ussinglehsize
\newskip\standardbaselineskip \newdimen\ussinglevsize
\newskip\reducedbaselineskip \newskip\doublebaselineskip
\eucolumnsize=12.0truecm    
\eudoublehsize=25.5truecm   
\eudoublevsize=6.5truein    
\uscolumnsize=4.4truein     
\usdoublehsize=9.4truein    
\usdoublevsize=6.8truein    
\eusinglehsize=6.5truein    
\eusinglevsize=24truecm     
\ussinglehsize=6.5truein    
\ussinglevsize=8.9truein    
\standardbaselineskip=16pt plus.2pt  
\reducedbaselineskip=14pt plus.2pt   
\doublebaselineskip=12pt plus.2pt    
%
%
\def\Portoffset{}
\def\Landoffset{}
\ifx\risposta\mplarisposta \def\Portoffset{\hoffset=1.8truecm} \fi
%
%
\def\Landspec{}
\tolerance=10000
\parskip=0pt plus2pt  \leftskip=0pt \rightskip=0pt
%
%
\ifx\risposta\standardrisposta \infralinea=\standardbaselineskip \fi
\ifx\risposta\reducedrisposta  \infralinea=\reducedbaselineskip \fi
\ifx\risposta\doublerisposta   \infralinea=\doublebaselineskip \fi
\ifx\risposta\mplarisposta     \infralinea=13pt \fi
\ifx\risposta\zerorisposta     \infralinea=12pt plus.2pt\fi
\ifnum\controllorisposta=0    \infralinea=\standardbaselineskip \fi
\ifx\risposta\doublerisposta   \Landoffset \else \Portoffset \fi
\ifx\risposta\doublerisposta \ifx\srisposta\cartarisposta
\tothsize=\eudoublehsize \collhsize=\eucolumnsize
\vsize=\eudoublevsize  \else  \tothsize=\usdoublehsize
\collhsize=\uscolumnsize \vsize=\usdoublevsize \fi \else
\ifx\srisposta\cartarisposta \tothsize=\eusinglehsize
\vsize=\eusinglevsize \else  \tothsize=\ussinglehsize
\vsize=\ussinglevsize \fi \collhsize=4.4truein \fi
\ifx\risposta\mplarisposta \tothsize=5.0truein
\vsize=7.8truein \collhsize=4.4truein \fi
%
%
%
%
\newcount\contaeuler \newcount\contacyrill \newcount\contaams
\font\ninerm=cmr9  \font\eightrm=cmr8  \font\sixrm=cmr6
\font\ninei=cmmi9  \font\eighti=cmmi8  \font\sixi=cmmi6
\font\ninesy=cmsy9  \font\eightsy=cmsy8  \font\sixsy=cmsy6
\font\ninebf=cmbx9  \font\eightbf=cmbx8  \font\sixbf=cmbx6
\font\ninett=cmtt9  \font\eighttt=cmtt8  \font\nineit=cmti9
\font\eightit=cmti8 \font\ninesl=cmsl9  \font\eightsl=cmsl8
\skewchar\ninei='177 \skewchar\eighti='177 \skewchar\sixi='177
\skewchar\ninesy='60 \skewchar\eightsy='60 \skewchar\sixsy='60
\hyphenchar\ninett=-1 \hyphenchar\eighttt=-1 \hyphenchar\tentt=-1
%
\font\tencmmib=cmmib10  \newfam\cmmibfam  \skewchar\tencmmib='177
\font\tencmbsy=cmbsy10  \newfam\cmbsyfam  \skewchar\tencmbsy='60
\def\scaps{\cmcsc}                 
\font\tencmcsc=cmcsc10  \newfam\cmcscfam
\ifnum\ingrandimento=1095

\font\capsone=cmcsc10 at 10.95pt 

\else

\font\capsone=cmcsc10 at 12pt 
\fi

\def\ttaarr{\bf}		
\def\ppaarr{\sl}		

%
%
%
\newfam\eufmfam \newfam\msamfam \newfam\msbmfam \newfam\eufbfam
\def\Loadeulerfonts{\global\contaeuler=1 \ifx\arisposta\amsrisposta
\font\teneufm=eufm10              
\font\eighteufm=eufm8 \font\nineeufm=eufm9 \font\sixeufm=eufm6
\font\seveneufm=eufm7  \font\fiveeufm=eufm5
\font\teneufb=eufb10              
\font\eighteufb=eufb8 \font\nineeufb=eufb9 \font\sixeufb=eufb6
\font\seveneufb=eufb7  \font\fiveeufb=eufb5
\font\teneurm=eurm10              
\font\eighteurm=eurm8 \font\nineeurm=eurm9
\font\teneurb=eurb10              
\font\eighteurb=eurb8 \font\nineeurb=eurb9
\font\teneusm=eusm10              
\font\eighteusm=eusm8 \font\nineeusm=eusm9
\font\teneusb=eusb10              
\font\eighteusb=eusb8 \font\nineeusb=eusb9
\else \def\eufm{\tt} \def\eufb{\tt} \def\eurm{\tt} \def\eurb{\tt}
\def\eusm{\tt} \def\eusb{\tt}    \fi}

\def\loadamsmath{\global\contaams=1 \ifx\arisposta\amsrisposta
\font\tenmsam=msam10 \font\ninemsam=msam9 \font\eightmsam=msam8
\font\sevenmsam=msam7 \font\sixmsam=msam6 \font\fivemsam=msam5
\font\tenmsbm=msbm10 \font\ninemsbm=msbm9 \font\eightmsbm=msbm8
\font\sevenmsbm=msbm7 \font\sixmsbm=msbm6 \font\fivemsbm=msbm5
\else \def\msbm{\bf} \fi \def\Bbb{\msbm} \def\symbl{\msam} \tenpoint}
\def\loadcyrill{\global\contacyrill=1 \ifx\arisposta\amsrisposta
\font\tenwncyr=wncyr10 \font\ninewncyr=wncyr9 \font\eightwncyr=wncyr8
\font\tenwncyb=wncyr10 \font\ninewncyb=wncyr9 \font\eightwncyb=wncyr8
\font\tenwncyi=wncyr10 \font\ninewncyi=wncyr9 \font\eightwncyi=wncyr8
\else \def\cyrill{\sl} \def\cyrilb{\sl} \def\cyrili{\sl} \fi\tenpoint}
\ifx\arisposta\amsrisposta
\font\sevenex=cmex7               
\font\eightex=cmex8  \font\nineex=cmex9
\font\ninecmmib=cmmib9   \font\eightcmmib=cmmib8
\font\sevencmmib=cmmib7 \font\sixcmmib=cmmib6
\font\fivecmmib=cmmib5   \skewchar\ninecmmib='177
\skewchar\eightcmmib='177  \skewchar\sevencmmib='177
\skewchar\sixcmmib='177   \skewchar\fivecmmib='177
\font\ninecmbsy=cmbsy9    \font\eightcmbsy=cmbsy8
\font\sevencmbsy=cmbsy7  \font\sixcmbsy=cmbsy6
\font\fivecmbsy=cmbsy5   \skewchar\ninecmbsy='60
\skewchar\eightcmbsy='60  \skewchar\sevencmbsy='60
\skewchar\sixcmbsy='60    \skewchar\fivecmbsy='60
\font\ninecmcsc=cmcsc9    \font\eightcmcsc=cmcsc8     \else
\def\cmmib{\fam\cmmibfam\tencmmib}\textfont\cmmibfam=\tencmmib
\scriptfont\cmmibfam=\tencmmib \scriptscriptfont\cmmibfam=\tencmmib
\def\cmbsy{\fam\cmbsyfam\tencmbsy} \textfont\cmbsyfam=\tencmbsy
\scriptfont\cmbsyfam=\tencmbsy \scriptscriptfont\cmbsyfam=\tencmbsy
\scriptfont\cmcscfam=\tencmcsc \scriptscriptfont\cmcscfam=\tencmcsc
\def\cmcsc{\fam\cmcscfam\tencmcsc} \textfont\cmcscfam=\tencmcsc \fi
\catcode`@=11
\newskip\ttglue
\gdef\tenpoint{\def\rm{\fam0\tenrm}
  \textfont0=\tenrm \scriptfont0=\sevenrm \scriptscriptfont0=\fiverm
  \textfont1=\teni \scriptfont1=\seveni \scriptscriptfont1=\fivei
  \textfont2=\tensy \scriptfont2=\sevensy \scriptscriptfont2=\fivesy
  \textfont3=\tenex \scriptfont3=\tenex \scriptscriptfont3=\tenex
  \def\mcal{\fam2 \tensy}  \def\mmit{\fam1 \teni}
  \textfont\itfam=\tenit \def\it{\fam\itfam\tenit}
  \textfont\slfam=\tensl \def\sl{\fam\slfam\tensl}
  \textfont\ttfam=\tentt \scriptfont\ttfam=\eighttt
  \scriptscriptfont\ttfam=\eighttt  \def\tt{\fam\ttfam\tentt}
  \textfont\bffam=\tenbf \scriptfont\bffam=\sevenbf
  \scriptscriptfont\bffam=\fivebf \def\bf{\fam\bffam\tenbf}
     \ifx\arisposta\amsrisposta    \ifnum\contaeuler=1
  \textfont\eufmfam=\teneufm \scriptfont\eufmfam=\seveneufm
  \scriptscriptfont\eufmfam=\fiveeufm \def\eufm{\fam\eufmfam\teneufm}
  \textfont\eufbfam=\teneufb \scriptfont\eufbfam=\seveneufb
  \scriptscriptfont\eufbfam=\fiveeufb \def\eufb{\fam\eufbfam\teneufb}
  \def\eurm{\teneurm} \def\eurb{\teneurb} \def\eusm{\teneusm}
  \def\eusb{\teneusb}    \fi    \ifnum\contaams=1
  \textfont\msamfam=\tenmsam \scriptfont\msamfam=\sevenmsam
  \scriptscriptfont\msamfam=\fivemsam \def\msam{\fam\msamfam\tenmsam}
  \textfont\msbmfam=\tenmsbm \scriptfont\msbmfam=\sevenmsbm
  \scriptscriptfont\msbmfam=\fivemsbm \def\msbm{\fam\msbmfam\tenmsbm}
     \fi      \ifnum\contacyrill=1     \def\cyrill{\tenwncyr}
  \def\cyrilb{\tenwncyb}  \def\cyrili{\tenwncyi}         \fi
  \textfont3=\tenex \scriptfont3=\sevenex \scriptscriptfont3=\sevenex
  \def\cmmib{\fam\cmmibfam\tencmmib} \scriptfont\cmmibfam=\sevencmmib
  \textfont\cmmibfam=\tencmmib  \scriptscriptfont\cmmibfam=\fivecmmib
  \def\cmbsy{\fam\cmbsyfam\tencmbsy} \scriptfont\cmbsyfam=\sevencmbsy
  \textfont\cmbsyfam=\tencmbsy  \scriptscriptfont\cmbsyfam=\fivecmbsy
  \def\cmcsc{\fam\cmcscfam\tencmcsc} \scriptfont\cmcscfam=\eightcmcsc
  \textfont\cmcscfam=\tencmcsc \scriptscriptfont\cmcscfam=\eightcmcsc
     \fi            \tt \ttglue=.5em plus.25em minus.15em
  \normalbaselineskip=12pt
  \setbox\strutbox=\hbox{\vrule height8.5pt depth3.5pt width0pt}
  \let\sc=\eightrm \let\big=\tenbig   \normalbaselines
  \baselineskip=\infralinea  \rm}
\gdef\ninepoint{\def\rm{\fam0\ninerm}
  \textfont0=\ninerm \scriptfont0=\sixrm \scriptscriptfont0=\fiverm
  \textfont1=\ninei \scriptfont1=\sixi \scriptscriptfont1=\fivei
  \textfont2=\ninesy \scriptfont2=\sixsy \scriptscriptfont2=\fivesy
  \textfont3=\tenex \scriptfont3=\tenex \scriptscriptfont3=\tenex
  \def\mcal{\fam2 \ninesy}  \def\mmit{\fam1 \ninei}
  \textfont\itfam=\nineit \def\it{\fam\itfam\nineit}
  \textfont\slfam=\ninesl \def\sl{\fam\slfam\ninesl}
  \textfont\ttfam=\ninett \scriptfont\ttfam=\eighttt
  \scriptscriptfont\ttfam=\eighttt \def\tt{\fam\ttfam\ninett}
  \textfont\bffam=\ninebf \scriptfont\bffam=\sixbf
  \scriptscriptfont\bffam=\fivebf \def\bf{\fam\bffam\ninebf}
     \ifx\arisposta\amsrisposta  \ifnum\contaeuler=1
  \textfont\eufmfam=\nineeufm \scriptfont\eufmfam=\sixeufm
  \scriptscriptfont\eufmfam=\fiveeufm \def\eufm{\fam\eufmfam\nineeufm}
  \textfont\eufbfam=\nineeufb \scriptfont\eufbfam=\sixeufb
  \scriptscriptfont\eufbfam=\fiveeufb \def\eufb{\fam\eufbfam\nineeufb}
  \def\eurm{\nineeurm} \def\eurb{\nineeurb} \def\eusm{\nineeusm}
  \def\eusb{\nineeusb}     \fi   \ifnum\contaams=1
  \textfont\msamfam=\ninemsam \scriptfont\msamfam=\sixmsam
  \scriptscriptfont\msamfam=\fivemsam \def\msam{\fam\msamfam\ninemsam}
  \textfont\msbmfam=\ninemsbm \scriptfont\msbmfam=\sixmsbm
  \scriptscriptfont\msbmfam=\fivemsbm \def\msbm{\fam\msbmfam\ninemsbm}
     \fi       \ifnum\contacyrill=1     \def\cyrill{\ninewncyr}
  \def\cyrilb{\ninewncyb}  \def\cyrili{\ninewncyi}         \fi
  \textfont3=\nineex \scriptfont3=\sevenex \scriptscriptfont3=\sevenex
  \def\cmmib{\fam\cmmibfam\ninecmmib}  \textfont\cmmibfam=\ninecmmib
  \scriptfont\cmmibfam=\sixcmmib \scriptscriptfont\cmmibfam=\fivecmmib
  \def\cmbsy{\fam\cmbsyfam\ninecmbsy}  \textfont\cmbsyfam=\ninecmbsy
  \scriptfont\cmbsyfam=\sixcmbsy \scriptscriptfont\cmbsyfam=\fivecmbsy
  \def\cmcsc{\fam\cmcscfam\ninecmcsc} \scriptfont\cmcscfam=\eightcmcsc
  \textfont\cmcscfam=\ninecmcsc \scriptscriptfont\cmcscfam=\eightcmcsc
     \fi            \tt \ttglue=.5em plus.25em minus.15em
  \normalbaselineskip=11pt
  \setbox\strutbox=\hbox{\vrule height8pt depth3pt width0pt}
  \let\sc=\sevenrm \let\big=\ninebig \normalbaselines\rm}
\gdef\eightpoint{\def\rm{\fam0\eightrm}
  \textfont0=\eightrm \scriptfont0=\sixrm \scriptscriptfont0=\fiverm
  \textfont1=\eighti \scriptfont1=\sixi \scriptscriptfont1=\fivei
  \textfont2=\eightsy \scriptfont2=\sixsy \scriptscriptfont2=\fivesy
  \textfont3=\tenex \scriptfont3=\tenex \scriptscriptfont3=\tenex
  \def\mcal{\fam2 \eightsy}  \def\mmit{\fam1 \eighti}
  \textfont\itfam=\eightit \def\it{\fam\itfam\eightit}
  \textfont\slfam=\eightsl \def\sl{\fam\slfam\eightsl}
  \textfont\ttfam=\eighttt \scriptfont\ttfam=\eighttt
  \scriptscriptfont\ttfam=\eighttt \def\tt{\fam\ttfam\eighttt}
  \textfont\bffam=\eightbf \scriptfont\bffam=\sixbf
  \scriptscriptfont\bffam=\fivebf \def\bf{\fam\bffam\eightbf}
     \ifx\arisposta\amsrisposta   \ifnum\contaeuler=1
  \textfont\eufmfam=\eighteufm \scriptfont\eufmfam=\sixeufm
  \scriptscriptfont\eufmfam=\fiveeufm \def\eufm{\fam\eufmfam\eighteufm}
  \textfont\eufbfam=\eighteufb \scriptfont\eufbfam=\sixeufb
  \scriptscriptfont\eufbfam=\fiveeufb \def\eufb{\fam\eufbfam\eighteufb}
  \def\eurm{\eighteurm} \def\eurb{\eighteurb} \def\eusm{\eighteusm}
  \def\eusb{\eighteusb}       \fi    \ifnum\contaams=1
  \textfont\msamfam=\eightmsam \scriptfont\msamfam=\sixmsam
  \scriptscriptfont\msamfam=\fivemsam \def\msam{\fam\msamfam\eightmsam}
  \textfont\msbmfam=\eightmsbm \scriptfont\msbmfam=\sixmsbm
  \scriptscriptfont\msbmfam=\fivemsbm \def\msbm{\fam\msbmfam\eightmsbm}
     \fi       \ifnum\contacyrill=1     \def\cyrill{\eightwncyr}
  \def\cyrilb{\eightwncyb}  \def\cyrili{\eightwncyi}         \fi
  \textfont3=\eightex \scriptfont3=\sevenex \scriptscriptfont3=\sevenex
  \def\cmmib{\fam\cmmibfam\eightcmmib}  \textfont\cmmibfam=\eightcmmib
  \scriptfont\cmmibfam=\sixcmmib \scriptscriptfont\cmmibfam=\fivecmmib
  \def\cmbsy{\fam\cmbsyfam\eightcmbsy}  \textfont\cmbsyfam=\eightcmbsy
  \scriptfont\cmbsyfam=\sixcmbsy \scriptscriptfont\cmbsyfam=\fivecmbsy
  \def\cmcsc{\fam\cmcscfam\eightcmcsc} \scriptfont\cmcscfam=\eightcmcsc
  \textfont\cmcscfam=\eightcmcsc \scriptscriptfont\cmcscfam=\eightcmcsc
     \fi             \tt \ttglue=.5em plus.25em minus.15em
  \normalbaselineskip=9pt
  \setbox\strutbox=\hbox{\vrule height7pt depth2pt width0pt}
  \let\sc=\sixrm \let\big=\eightbig \normalbaselines\rm }
\gdef\tenbig#1{{\hbox{$\left#1\vbox to8.5pt{}\right.\n@space$}}}
\gdef\ninebig#1{{\hbox{$\textfont0=\tenrm\textfont2=\tensy
   \left#1\vbox to7.25pt{}\right.\n@space$}}}
\gdef\eightbig#1{{\hbox{$\textfont0=\ninerm\textfont2=\ninesy
   \left#1\vbox to6.5pt{}\right.\n@space$}}}
\def\alternativefont#1#2{\ifx\arisposta\amsrisposta \relax \else
\xdef#1{#2} \fi}
\global\contaeuler=0 \global\contacyrill=0 \global\contaams=0
%
%
%
%
\newbox\fotlinebb \newbox\hedlinebb \newbox\leftcolumn
\gdef\makeheadline{\vbox to 0pt{\vskip-22.5pt
     \fullline{\vbox to8.5pt{}\the\headline}\vss}\nointerlineskip}
\gdef\makehedlinebb{\vbox to 0pt{\vskip-22.5pt
     \fullline{\vbox to8.5pt{}\copy\hedlinebb\hfil
     \line{\hfill\the\headline\hfill}}\vss} \nointerlineskip}
\gdef\makefootline{\baselineskip=24pt \fullline{\the\footline}}
\gdef\makefotlinebb{\baselineskip=24pt
    \fullline{\copy\fotlinebb\hfil\line{\hfill\the\footline\hfill}}}
\gdef\doubleformat{\shipout\vbox{\Landspec\makehedlinebb
     \fullline{\box\leftcolumn\hfil\columnbox}\makefotlinebb}
     \advancepageno}
\gdef\columnbox{\leftline{\pagebody}}
\gdef\line#1{\hbox to\hsize{\hskip\leftskip#1\hskip\rightskip}}
\gdef\fullline#1{\hbox to\fullhsize{\hskip\leftskip{#1}%
\hskip\rightskip}}
\gdef\footnote#1{\let\@sf=\empty
         \ifhmode\edef\#sf{\spacefactor=\the\spacefactor}\/\fi
         #1\@sf\vfootnote{#1}}
\gdef\vfootnote#1{\insert\footins\bgroup
         \ifnum\dimnota=1  \eightpoint\fi
         \ifnum\dimnota=2  \ninepoint\fi
         \ifnum\dimnota=0  \tenpoint\fi
         \interlinepenalty=\interfootnotelinepenalty
         \splittopskip=\ht\strutbox
         \splitmaxdepth=\dp\strutbox \floatingpenalty=20000
         \leftskip=\oldssposta \rightskip=\olddsposta
         \spaceskip=0pt \xspaceskip=0pt
         \ifnum\sinnota=0   \textindent{#1}\fi
         \ifnum\sinnota=1   \item{#1}\fi
         \footstrut\futurelet\next\fo@t}
\gdef\fo@t{\ifcat\bgroup\noexpand\next \let\next\f@@t
             \else\let\next\f@t\fi \next}
\gdef\f@@t{\bgroup\aftergroup\@foot\let\next}
\gdef\f@t#1{#1\@foot} \gdef\@foot{\strut\egroup}
\gdef\footstrut{\vbox to\splittopskip{}}
\skip\footins=\bigskipamount
\count\footins=1000  \dimen\footins=8in
\catcode`@=12
\tenpoint
\ifnum\unoduecol=1 \hsize=\tothsize   \fullhsize=\tothsize \fi
\ifnum\unoduecol=2 \hsize=\collhsize  \fullhsize=\tothsize \fi
\global\let\lrcol=L      \ifnum\unoduecol=1
\output{\plainoutput{\ifnum\tipbnota=2 \clearnmbnota\fi}} \fi
\ifnum\unoduecol=2 \output{\if L\lrcol
     \global\setbox\leftcolumn=\columnbox
     \global\setbox\fotlinebb=\line{\hfill\the\footline\hfill}
     \global\setbox\hedlinebb=\line{\hfill\the\headline\hfill}
     \advancepageno  \global\let\lrcol=R
     \else  \doubleformat \global\let\lrcol=L \fi
     \ifnum\outputpenalty>-20000 \else\dosupereject\fi
     \ifnum\tipbnota=2\clearnmbnota\fi }\fi
\def\ifdoublepage{\ifnum\unoduecol=2 }
\gdef\yespagenumbers{\footline={\hss\tenrm\folio\hss}}
\gdef\ciao{ \ifnum\fdefcontre=1 \endfdef\fi
     \par\vfill\supereject \ifnum\unoduecol=2
     \if R\lrcol  \headline={}\nopagenumbers\null\vfill\eject
     \fi\fi \end}

\newskip\olddsposta \newskip\oldssposta
\global\oldssposta=\leftskip \global\olddsposta=\rightskip

\def\filldots{\leaders\hbox to 1em{\hss.\hss}\hfill}
\def\inquadrb#1 {\vbox {\hrule  \hbox{\vrule \vbox {\vskip .2cm
    \hbox {\ #1\ } \vskip .2cm } \vrule  }  \hrule} }
 \def\newline{\hfil\break}
\def\jump{\vskip\baselineskip} \newskip\iinnffrr
\def\sjump{\iinnffrr=\baselineskip
          \divide\iinnffrr by 2 \vskip\iinnffrr}
\def\bjump{\vskip\baselineskip \vskip\baselineskip}
\newcount\nmbnota  \def\clearnmbnota{\global\nmbnota=0}
\newcount\tipbnota \def\letterfootnote{\global\tipbnota=1}

\def\note#1{\global\advance\nmbnota by 1 \ifnum\tipbnota=1
    \footnote{$^{\rm\nttlett}$}{#1} \else {\ifnum\tipbnota=2
    \footnote{$^{\nttsymb}$}{#1}
    \else\footnote{$^{\the\nmbnota}$}{#1}\fi}\fi}
\def\nttlett{\ifcase\nmbnota \or a\or b\or c\or d\or e\or f\or
g\or h\or i\or j\or k\or l\or m\or n\or o\or p\or q\or r\or
s\or t\or u\or v\or w\or y\or x\or z\fi}
\def\nttsymb{\ifcase\nmbnota \or\dag\or\sharp\or\ddag\or\star\or
\natural\or\flat\or\clubsuit\or\diamondsuit\or\heartsuit
\or\spadesuit\fi}   \clearnmbnota
\def\numberfootnote{\global\tipbnota=0} \numberfootnote
\def\setnote#1{\expandafter\xdef\csname#1\endcsname{
\ifnum\tipbnota=1 {\rm\nttlett} \else {\ifnum\tipbnota=2
{\nttsymb} \else \the\nmbnota\fi}\fi} }
\newcount\nbmfig  \def\clearnbmfig{\global\nbmfig=0}
\gdef\figure{\global\advance\nbmfig by 1
      {\rm fig. \the\nbmfig}}   \clearnbmfig
\def\setfig#1{\expandafter\xdef\csname#1\endcsname{fig. \the\nbmfig}}
 \def\endformula{\eqno\numero $$}
 \def\efr{\endformula}
\newcount\frmcount \def\clearfrmcount{\global\frmcount=0}
\def\numero{\global\advance\frmcount by 1   \ifnum\indappcount=0
  {\ifnum\cpcount <1 {\hbox{\rm (\the\frmcount )}}  \else
  {\hbox{\rm (\the\cpcount .\the\frmcount )}} \fi}  \else
  {\hbox{\rm (\applett .\the\frmcount )}} \fi}
\def\nameformula#1{\global\advance\frmcount by 1%
\ifnum\draftnum=0  {\ifnum\indappcount=0%
{\ifnum\cpcount<1\xdef\spzzttrra{(\the\frmcount )}%
\else\xdef\spzzttrra{(\the\cpcount .\the\frmcount )}\fi}%
\else\xdef\spzzttrra{(\applett .\the\frmcount )}\fi}%
\else\xdef\spzzttrra{(#1)}\fi%
\expandafter\xdef\csname#1\endcsname{\spzzttrra}
\eqno \hbox{\rm\spzzttrra} $$}
\def\nfr{\nameformula}    
\def\nameali#1{\global\advance\frmcount by 1%
\ifnum\draftnum=0  {\ifnum\indappcount=0%
{\ifnum\cpcount<1\xdef\spzzttrra{(\the\frmcount )}%
\else\xdef\spzzttrra{(\the\cpcount .\the\frmcount )}\fi}%
\else\xdef\spzzttrra{(\applett .\the\frmcount )}\fi}%
\else\xdef\spzzttrra{(#1)}\fi%
\expandafter\xdef\csname#1\endcsname{\spzzttrra}
  \hbox{\rm\spzzttrra} }      \clearfrmcount
\newcount\cpcount \def\clearcpcount{\global\cpcount=0}
\newcount\subcpcount \def\clearsubcpcount{\global\subcpcount=0}
\newcount\appcount \def\clearappcount{\global\appcount=0}
\newcount\indappcount \def\clearindappcount{\indappcount=0}
\newcount\sottoparcount 

\def\applett{\ifcase\appcount  \or {A}\or {B}\or {C}\or
{D}\or {E}\or {F}\or {G}\or {H}\or {I}\or {J}\or {K}\or {L}\or
{M}\or {N}\or {O}\or {P}\or {Q}\or {R}\or {S}\or {T}\or {U}\or
{V}\or {W}\or {X}\or {Y}\or {Z}\fi    \ifnum\appcount<0
\immediate\write16 {Panda ERROR - Appendix: counter "appcount"
out of range}\fi  \ifnum\appcount>26  \immediate\write16 {Panda
ERROR - Appendix: counter "appcount" out of range}\fi}
\clearappcount  \clearindappcount \newcount\connttrre
\def\clearconnttrre{\global\connttrre=0} \newcount\countref
\def\clearcountref{\global\countref=0} \clearcountref
\def\chapter#1{\global\advance\cpcount by 1 \clearfrmcount
                 \goodbreak\null\vbox{\jump\nobreak
                 \clearsubcpcount\clearindappcount
                 \itemitem{\ttaarr\the\cpcount .\qquad}{\ttaarr #1}
                 \par\nobreak\jump\sjump}\nobreak}
\def\section#1{\global\advance\subcpcount by 1 \goodbreak\null
               \vbox{\sjump\nobreak\ifnum\indappcount=0
                 {\ifnum\cpcount=0 {\itemitem{\ppaarr
               .\the\subcpcount\quad\enskip\ }{\ppaarr #1}\par} \else
                 {\itemitem{\ppaarr\the\cpcount .\the\subcpcount\quad
                  \enskip\ }{\ppaarr #1} \par}  \fi}
                \else{\itemitem{\ppaarr\applett .\the\subcpcount\quad
                 \enskip\ }{\ppaarr #1}\par}\fi\nobreak\jump}\nobreak}
\clearsubcpcount
\def\appendix#1{\global\advance\appcount by 1 \clearfrmcount
                  \goodbreak\null\vbox{\jump\nobreak
                  \global\advance\indappcount by 1 \clearsubcpcount
          \itemitem{ }{\hskip-40pt\ttaarr Appendix\ \applett :\ #1}
             \nobreak\jump\sjump}\nobreak}
\clearappcount \clearindappcount
\def\references{\goodbreak\null\vbox{\jump\nobreak
   \itemitem{}{\ttaarr References} \nobreak\jump\sjump}\nobreak}

\clearcpcount\clearcountref

\def\setchap#1{\ifnum\indappcount=0{\ifnum\subcpcount=0%
\xdef\spzzttrra{\the\cpcount}%
\else\xdef\spzzttrra{\the\cpcount .\the\subcpcount}\fi}
\else{\ifnum\subcpcount=0 \xdef\spzzttrra{\applett}%
\else\xdef\spzzttrra{\applett .\the\subcpcount}\fi}\fi
\expandafter\xdef\csname#1\endcsname{\spzzttrra}}
\newcount\draftnum \newcount\ppora   \newcount\ppminuti
\global\ppora=\time   \global\ppminuti=\time
\global\divide\ppora by 60  \draftnum=\ppora
\multiply\draftnum by 60    \global\advance\ppminuti by -\draftnum
\def\droggi{\number\day /\number\month /\number\year\ \the\ppora
:\the\ppminuti}     \global\draftnum=0
\def\draftcomment#1{\ifnum\draftnum=0 \relax \else
{\ {\bf ***}\ #1\ {\bf ***}\ }\fi} 
%
%
\catcode`@=11
\gdef\Ref#1{\expandafter\ifx\csname @rrxx@#1\endcsname\relax%
{\global\advance\countref by 1    \ifnum\countref>200
\immediate\write16 {Panda ERROR - Ref: maximum number of references
exceeded}  \expandafter\xdef\csname @rrxx@#1\endcsname{0}\else
\expandafter\xdef\csname @rrxx@#1\endcsname{\the\countref}\fi}\fi
\ifnum\draftnum=0 \csname @rrxx@#1\endcsname \else#1\fi}
\gdef\beginref{\ifnum\draftnum=0  \gdef\Rref{\fairef}
\gdef\endref{\scriviref} \else\relax\fi
\ifx\risposta\mplarisposta \ninepoint \fi
\parskip 2pt plus.2pt \baselineskip=12pt}
\def\Reflab#1{[#1]} \gdef\Rref#1#2{\item{\Reflab{#1}}{#2}}
\gdef\endref{\relax}  \newcount\conttemp
\gdef\fairef#1#2{\expandafter\ifx\csname @rrxx@#1\endcsname\relax
{\global\conttemp=0 \immediate\write16 {Panda ERROR - Ref: reference
[#1] undefined}} \else
{\global\conttemp=\csname @rrxx@#1\endcsname } \fi
\global\advance\conttemp by 50  \global\setbox\conttemp=\hbox{#2} }
\gdef\scriviref{\clearconnttrre\conttemp=50
\loop\ifnum\connttrre<\countref \advance\conttemp by 1
\advance\connttrre by 1
\item{\Reflab{\the\connttrre}}{\unhcopy\conttemp} \repeat}
\clearcountref \clearconnttrre
\catcode`@=12
\ifx\risposta\mplarisposta \def\Reflab#1{#1.} \letterfootnote \fi

\def\slashchar#1{\setbox0=\hbox{$#1$} \dimen0=\wd0
     \setbox1=\hbox{/} \dimen1=\wd1 \ifdim\dimen0>\dimen1
      \rlap{\hbox to \dimen0{\hfil/\hfil}} #1 \else
      \rlap{\hbox to \dimen1{\hfil$#1$\hfil}} / \fi}
\ifx\oldchi\undefined \let\oldchi=\chi
  \def\cchi{{\raise 1pt\hbox{$\oldchi$}}} \let\chi=\cchi \fi

\def\frac#1#2{{\textstyle{#1 \over #2}}}

\def\half{\ifinner {\scriptstyle {1 \over 2}}\else {1 \over 2} \fi}

\def\simge{\rlap{\raise 2pt \hbox{$>$}}{\lower 2pt \hbox{$\sim$}}}
\def\simle{\rlap{\raise 2pt \hbox{$<$}}{\lower 2pt \hbox{$\sim$}}}

\def\vbig#1#2{{\vbigd@men=#2\divide\vbigd@men by 2%
\hbox{$\left#1\vbox to \vbigd@men{}\right.\n@space$}}}

%
%
\newcount\fdefcontre \newcount\fdefcount \newcount\indcount
\newread\filefdef  \newread\fileftmp  \newwrite\filefdef
\newwrite\fileftmp     \def\strip#1*.A {#1}
\def\futuredef#1{\beginfdef
\expandafter\ifx\csname#1\endcsname\relax%
{\immediate\write\fileftmp {#1*.A}
\immediate\write16 {Panda Warning - fdef: macro "#1" on page
\the\pageno \space undefined}
\ifnum\draftnum=0 \expandafter\xdef\csname#1\endcsname{(?)}
\else \expandafter\xdef\csname#1\endcsname{(#1)} \fi
\global\advance\fdefcount by 1}\fi   \csname#1\endcsname}

\def\beginfdef{\ifnum\fdefcontre=0
\immediate\openin\filefdef \jobname.fdef
\immediate\openout\fileftmp \jobname.ftmp
\global\fdefcontre=1  \ifeof\filefdef \immediate\write16 {Panda
WARNING - fdef: file \jobname.fdef not found, run TeX again}
\else \immediate\read\filefdef to\spzzttrra
\global\advance\fdefcount by \spzzttrra
\indcount=0      \loop\ifnum\indcount<\fdefcount
\advance\indcount by 1   \immediate\read\filefdef to\spezttrra
\immediate\read\filefdef to\sppzttrra
\edef\spzzttrra{\expandafter\strip\spezttrra}
\immediate\write\fileftmp {\spzzttrra *.A}
\expandafter\xdef\csname\spzzttrra\endcsname{\sppzttrra}
\repeat \fi \immediate\closein\filefdef \fi}
\def\endfdef{\immediate\closeout\fileftmp   \ifnum\fdefcount>0
\immediate\openin\fileftmp \jobname.ftmp
\immediate\openout\filefdef \jobname.fdef
\immediate\write\filefdef {\the\fdefcount}   \indcount=0
\loop\ifnum\indcount<\fdefcount    \advance\indcount by 1
\immediate\read\fileftmp to\spezttrra
\edef\spzzttrra{\expandafter\strip\spezttrra}
\immediate\write\filefdef{\spzzttrra *.A}
\edef\spezttrra{\string{\csname\spzzttrra\endcsname\string}}
\iwritel\filefdef{\spezttrra}
\repeat  \immediate\closein\fileftmp \immediate\closeout\filefdef
\immediate\write16 {Panda Warning - fdef: Label(s) may have changed,
re-run TeX to get them right}\fi}
\def\iwritel#1#2{\newlinechar=-1
{\newlinechar=`\ \immediate\write#1{#2}}\newlinechar=-1}
\global\fdefcontre=0 \global\fdefcount=0 \global\indcount=0
%
%
\null
%
%
%
%

\input psfig
%
\loadamsmath
%
\pageno=0\baselineskip=14pt
\nopagenumbers{
\line{\hfill CERN-TH.6888/93}
\line{\hfill\tt hep-th/9305042}
\line{\hfill May 1993}
\ifdoublepage \bjump\bjump\bjump\bjump\else\vfill\fi
\centerline{\capsone THE ANALYTIC STRUCTURE OF}
\sjump\sjump
\centerline{\capsone TRIGONOMETRIC S MATRICES}
\bjump\bjump
\centerline{\scaps Timothy J. Hollowood}
\sjump
\sjump
\centerline{\sl CERN-TH, 1211 Geneva 23, Switzerland.}
\centerline{\tt hollow@surya11.cern.ch}
\bjump\bjump\bjump
\ifdoublepage
\vfill
\noindent
\line{CERN-TH.6888/93\hfill}
\line{May 1993\hfill}
\eject\null\vfill\fi
\centerline{\capsone ABSTRACT}\sjump
$S$-matrices associated to the vector representations of the
quantum groups for the classical Lie algebras are constructed. For the
$a_{m-1}$ and $c_m$ algebras the complete $S$-matrix is found by an
application of the bootstrap equations. It is shown that the simplest
form for the $S$-matrix which generalizes that of the Gross-Neveu model
is not consistent for the non-simply-laced algebras due to the existence
of unexplained singularities on the physical strip. However, a form which
generalizes the $S$-matrix of the principal chiral model is shown to be
consistent via an argument which uses a novel application of the
Coleman-Thun mechanism. The analysis also gives a correct description of the
analytic structure of the $S$-matrix of the principle chiral model for $c_m$.
\sjump\vfill
\ifdoublepage \else
\noindent
\line{CERN-TH.6888/93\hfill}
\line{May 1993\hfill}\fi
\eject}
\yespagenumbers\pageno=1
%
%

\chapter{Introduction}

The complete non-perturbative solution of interacting
quantum field theories seems like a hopeless fantasy. In the vast space
of two-dimensional quantum field theories, however, there is a small
set of theories attracting a large amount of interest due to the fact
that they are integrable and hence to a certain degree solvable.
At the moment
the understanding of such theories is mostly at the level of on-shell
physics encoded in the scattering matrix.
Even though the $S$-matrix seems to be rather simple, since it factorizes
and is specified completely by the two-body $S$-matrix, finding the complete
$S$-matrix satisfying all the required properties is a surprisingly difficult
task. Also it is worth pointing out that a given $S$-matrix
may always be multiplied by CDD factors to give an equivalent $S$-matrix.

Let us summarize the situation so far. First of all, there exist a
set of minimal purely elastic $S$-matrices (and hence trivial solutions of the
Yang-Baxter equation) which are related to the simply-laced
affine Lie algebras. There are particular CDD factors, which depend on a
coupling constant which make these
$S$-matrices equal to the conjectured $S$-matrices of the corresponding
affine Toda field theory (the connection with the field theory being
established in perturbation theory) [\Ref{TOD}]. There are also a set of
$S$-matrices
which describe the non-simply-laced affine Toda field theories, however, the
situation is rather different from the simply-laced theories in that the
$S$-matrix does not factor into a `minimal' piece independent of the
coupling and furthermore the mass ratios depend on the coupling constant
[\Ref{NLS}].

We now turn to the $S$-matrices that are non-trivial solutions of
the Yang-Baxter equation, so that the particle states carry some internal
quantum
numbers. One has both the rational and trigonometric solutions of the
Yang-Baxter equation to hand, the latter depending on a coupling constant.
$S$-matrices corresponding to rational solutions
have been considered in [\Ref{BOOT}], and describe the
principal chiral model for a classical Lie algebra
and the Gross-Neveu model for the simply-laced classical Lie algebras.
It is worth pointing out that a Gross-Neveu type $S$-matrix for
the non-simply-laced algebras can be written down, but the Ansatz fails
due to the existence of unexplained singularities on the physical strip.
This point highlights the
delicate nature of these $S$-matrices; it is simply not good enough to write
down an Ansatz for the $S$-matrix elements on some elementary particles,
hoping that the bootstrap will close on a conjectured set of
particles; in order to claim a consistent $S$-matrix it is necessary
to account for all singularities on the physical strip in terms of bound
states or anomalous thresholds via the Coleman-Thun mechanism [\Ref{OSD}].

Trigonometric solutions of the Yang-Baxter equation
are more general since they depend on a coupling constant and the rational
solutions can be obtained from them in a certain limit.
$S$ matrices constructed from
trigonometric solutions have been the subject of much speculation but
very little detailed analysis, except for the series associated to $a_{m-1}$
[\Ref{MSQ},\Ref{VF}] and in particular for $a_1$ which gives the soliton
$S$-matrix of the sine-Gordon theory [\Ref{ZZ}].
In these cases one can prove that the bootstrap closes on a given set of
particles, when the coupling constant satisfies a certain inequality
[\Ref{MSQ}]. The problem encountered for other algebras stems from
the difficulties in solving the bootstrap equations. However, we shall show
how this can be achieved for $c_m$ (in addition to $a_{m-1}$ which has been
considered previously).

One of the principle reasons for constructing the trigonometric $S$-matrices
is that they are thought to describe the integrable perturbations of
certain conformal field theories, typically displaying $W$-algebra type
symmetries [\Ref{VF},\Ref{PERT}].

\chapter{Trigonometric solutions of the Yang-Baxter equation}

\def\rc{\check R}
\def\ot{\otimes}

In this section we describe the trigonometric solutions of the Yang-Baxter
Equation (YBE). They are associated to certain deformations of
the universal enveloping algebra of a
Lie algebra known as a quantum group [\Ref{JIM},\Ref{DRIN}].
The solutions can be thought of as intertwiners between
tensor products of representations of the algebra:
$$
\rc(u):\ V_\mu\otimes V_\nu\rightarrow V_\nu\otimes V_\mu,
\nfr{INT}
where $u$ is the (additive) spectral parameter. Such an `$R$-matrix'
has a spectral decomposition
$$
\rc(u)=\sum_{\lambda}\rho_\lambda(u){\Bbb P}_\lambda,
\efr
where ${\Bbb P}_\lambda$ is a quantum group invariant homomorphism
$V_\mu\otimes V_\nu\rightarrow V_\nu\otimes V_\mu$ with the property that
${\Bbb P}_\lambda|_{V_{\lambda'}}\neq0$ if and only if $\lambda'=\lambda$.
\note{We are assuming that every irreducible
component $V_\lambda\subset V_\mu\otimes V_\nu$ has multiplicity one.}
If $\mu=\nu$ then ${\Bbb P}_\lambda$ is a projection.

In the following we shall use both the language of spectral decompositions
and the interaction-round-a-face (IRF) picture when writing down solutions
of the YBE. When using the latter we shall denote $\rc(u)$ as
$W(u)$. To start with we consider
the solutions associated to the vector representation of the algebra. In
what follows we shall consider all the classical Lie algebras: $a_{m-1}$,
$b_m$, $c_m$ and $d_m$. The set of weights $\Sigma$ of the vector
representations are\note{In our
notation the long roots of $b_m$ have length 2 while for $c_m$ the single
long root has length 4.}
$$\eqalign{
&\Sigma=\left\{e_1-(e_1+\cdots+e_m)/m,\ldots,e_m-(e_1+\cdots
+e_m)/m\right\},\quad {\rm for}\ a_{m-1},\cr
&\Sigma=\left\{0,\pm e_1,\ldots,\pm e_m\right\},\quad {\rm for}\ b_m,
\qquad\Sigma=\left\{\pm e_1,\ldots,\pm e_m\right\},\quad {\rm for}\ c_m,
d_m,\cr}
\efr
where the $e_i$'s are a set of orthonormal vectors.

The solution of the YBE is labelled by four weights of the algebra:
$$
W\left(\left.\matrix{a&b\cr c&d\cr}\right\vert u\right),\quad a,b,c,d\in
\Lambda^\star,
\efr
with the property that $W$ is only non-zero if $c-a$, $d-c$, $b-a$ and $d-b$
are $\in\Sigma$.

For completeness we now write down the solutions following [\Ref{JMO}] (see
also the review [\Ref{WDA}]). In the following $\omega$ is a constant which
is related to the deformation parameter of the quantum group.
For convenience we introduce for $a\in\Lambda^\star$
$$\eqalign{
&a_\mu=\omega(a+\rho)\cdot\mu,\quad{\rm for}\ \mu\in\Sigma\neq0,\cr
&a_0=-\omega/2,\quad a_{\mu\nu}=a_\mu-a_\nu,\quad
a_{\mu-\nu}=a_\mu+a_\nu,\cr}
\efr
where $\rho$ is the sum of the fundamental weights of the algebra.\note{These
are the vectors $\omega_i$ with $\omega_i\cdot\alpha_j=(\alpha_j^2/2)\delta_
{ij}$ where the $\alpha_j$ are the simple roots.} Also we define
$$
[x]=\sin x,\qquad\lambda=tg\omega/2,
\efr
where $g$ is the dual Coxeter number of the algebra and $t$ is the
$({\rm length})^2/2$ of the longest root.\note{The values $(g,t)$ are $(m,1)$,
$(2m-1,1)$, $(m+1,2)$ and $(2m-2,1)$ for $a_{m-1}$, $b_m$, $c_m$ and $d_m$,
respectively.}

For $a_{m-1}$ the solution is
$$\eqalign{
&W\left(\left.\matrix{a&a+\mu\cr a+\mu&a+2\mu\cr}\right\vert u\right)=
[\omega-\lambda u]/[\omega],\cr
&W\left(\left.\matrix{a&a+\mu\cr a+\mu&a+\mu+\nu\cr}\right\vert u\right)=
[a_{\mu\nu}+\lambda u]/[a_{\mu\nu}],\cr
&W\left(\left.\matrix{a&a+\nu\cr a+\mu&a+\mu+\nu\cr}\right\vert u\right)=
{[\lambda u]\over[\omega]}\left({[a_{\mu\nu}+\omega][a_{\mu\nu}-\omega]\over
[a_{\mu\nu}]^2}\right)^{1/2},\cr}
\nfr{SA}
where $\mu$, $\nu\in\Sigma$ and $\mu\neq\nu$. For the algebras $b_m$, $c_m$
and $d_m$ the solutions are
$$\eqalign{
&W\left(\left.\matrix{a&a+\mu\cr a+\mu&a+2\mu\cr}\right\vert u\right)=
{[\lambda-\lambda u][\omega-\lambda u]\over[\lambda][\omega]},
\quad{\rm for}\ \mu\neq0,\cr
&W\left(\left.\matrix{a&a+\mu\cr a+\mu&a+\mu+\nu\cr}\right\vert u\right)=
{[\lambda-\lambda u][a_{\mu\nu}+\lambda u]\over[\lambda]
[a_{\mu\nu}]},\quad{\rm for}\ \mu\neq\pm\nu,\cr
&W\left(\left.\matrix{a&a+\nu\cr a+\mu&a+\mu+\nu\cr}\right\vert u\right)=
{[\lambda-\lambda u][\lambda u]\over[\lambda][\omega]}\cr
&\qquad\qquad\qquad\qquad\qquad\qquad
\times\left({[a_{\mu\nu}+\omega][a_{\mu\nu}-
\omega]\over[a_{\mu\nu}]^2}\right)^{1/2},\quad{\rm for}\ \mu\neq\pm\nu,\cr
&W\left(\left.\matrix{a&a+\nu\cr a+\mu&a\cr}\right\vert u\right)=
{[\lambda u][a_{\mu-\nu}+\omega-\lambda+\lambda u]\over
[\lambda][a_{\mu-\nu}+\omega]}
\left(G_{a_\mu}G_{a_\nu}\right)^{1/2}\cr
&\qquad\qquad\qquad\qquad\qquad\qquad+\delta_{\mu\nu}
{[\lambda-\lambda u][a_{\mu-\nu}+\omega+\lambda u]\over
[\lambda][a_{\mu-\nu}+\omega]},\quad{\rm for}\ \mu\neq0,\cr
&W\left(\left.\matrix{a&a\cr a&a\cr}\right\vert u\right)={[\lambda+\lambda
u][2\lambda
-\lambda u]\over[\lambda][2\lambda]}-{[\lambda u][\lambda-\lambda u]
\over[\lambda][2\lambda]}J_a,\cr}
\nfr{SBCD}
where $\mu$, $\nu\in\Sigma$ and
$$\eqalign{
&G_{a_\mu}=\sigma{s(a_\mu+\omega)\over s(a_\mu)}\prod_{\kappa\neq\pm\mu,0}
{[a_{\mu\kappa}+\omega]\over[a_{\mu\kappa}]},\quad{\rm for}\
\mu\neq0,\quad G_{a_0}=1,\cr
&J_a=\sum_{\kappa\neq0}{[a_\kappa+\omega/2-2\lambda]\over[a_\kappa+
\omega/2]}G_{a_{\kappa}}.\cr}
\nfr{GJ}
In the above $\sigma$ is $-1$ for $c_m$ otherwise being $1$. The function
$s(x)=[tx]$ for $b_m$ and $c_m$, and $s(x)=1$ for $d_m$.

The solutions satisfy a set of conditions, in addition to the YBE,
which will be important for the construction of an $S$-matrix (our
terminology follows that of [\Ref{WDA}]).

(i) The standard initial condition.
$$
W\left(\left.\matrix{a&b\cr c&d\cr}\right\vert 0\right)=\delta_{bc}.
\nfr{IC}

(ii) The unitarity condition.
$$
\sum_eW\left(\left.\matrix{a&e\cr c&d\cr}\right\vert u\right)
W\left(\left.\matrix{a&b\cr e&d\cr}\right\vert -u\right)=\varrho(u)
\delta_{bc},
\nfr{CR}
where
$$\eqalign{
\varrho(u)=&{[\omega-\lambda u][\omega+\lambda u]\over[\omega]^2},
\quad{\rm for}\ a_{m-1},\cr
=&{[\lambda-\lambda u][\omega-\lambda u][\lambda+\lambda u][\omega+\lambda u]
\over[\lambda]^2[\omega]^2},\quad{\rm for}\ b_m,\ c_m\ {\rm and}\ d_m.\cr}
\nfr{RHOD}

(iii) Crossing symmetry ($b_m$, $c_m$ and $d_m$ only).
$$
W\left(\left.\matrix{a&b\cr c&d\cr}\right\vert u\right)=
W\left(\left.\matrix{c&a\cr d&b\cr}\right\vert 1-u\right)
\left({G_bG_c\over G_aG_d}\right)^{1/2},
\nfr{CS}
where
$$
G_a=\varepsilon(a)\prod_{k=1}^ms(a_i)\prod_{1\leq i<j\leq m}
[a_i-a_j][a_i+a_j],
\efr
where $a=\sum_{i=1}^ma_ie_i$ defines the $a_i$
and $\varepsilon(a)$ is a sign factor chosen
so that $\varepsilon(a+\mu)/\varepsilon(a)=\sigma$. $G_a$ is related to
$G_{a_\mu}$ in \GJ\ by $G_{a_\mu}=G_{a+\mu}/G_a$.

For $a_{m-1}$ the representation is complex and so the solution
of the YBE does not satisfy a crossing symmetry relation involving only
the vector representation.

The above solutions correspond to the unrestricted solutions of the YBE
where the variables $\{a,b,c,d\}$ are any weights of the algebra.
We shall be primarily interested in the restricted models which are obtained
for the particular values
$$
\omega={\pi\over t(g+k)},\quad k=1,2,\ldots
\nfr{RESM}
and the weights are restricted to lie in the set of integrable weights
of the affine algebra at level $k$ projected onto the weight lattice of the
finite algebra. In practice, this means that we restrict the allowed weights
to the dominant weights satisfying
$$
a\cdot\theta\leq k,
\efr
where $\theta$ is the highest root of the algebra. We denote the set of
weights satisfying this condition as $\Lambda^\star(k)$.

\chapter{The vector-vector scattering matrices}

The solutions of the YBE equation that we wrote down in the last section
naturally lead to $S$-matrices for a set of kinks. We denote a kink state by
$K_{ab}(\theta)$, where $a$ and $b$ are two vacua of the theory and $\theta$
is the rapidity of the kink. In an integrable field theory we need only
consider the $S$-matrix for the process
$$
K_{ac}(\theta_1)+K_{cd}(\theta_2)\rightarrow K_{ab}(\theta_2)+K_{bd}
(\theta_1)
\nfr{KP}
since all the other $S$-matrix elements are determined in terms of these.
The idea is to find an $S$-matrix for a theory associated to a
classical Lie algebra with vacua in one-to-one correspondence with
$\Lambda^\star(k)$ (or $\Lambda^\star$ in the unrestricted model) and
hence kinks with a topological charge being a weight of the vector
representation of the algebra.
The $S$-matrix of the process \KP\ is related to the solution of the YBE as
$$
{\widetilde S}\left(\left.\matrix{a&b\cr c&d\cr}\right\vert u\right)=
Y(u)
W\left(\left.\matrix{a&b\cr c&d\cr}\right\vert f(u)\right)\left({G_aG_d\over
G_bG_c}\right)^{f(u)/2},
\nfr{SMAT}
for some scalar function $Y(u)$, where $u=(\theta_
1-\theta_2)/i\pi$ is the rapidity difference of the incoming kinks.
The other
scalar factor is included in order to simplify the implementation of crossing
symmetry. The scalar function
$Y(u)$ must be chosen so that the $S$-matrix axioms are satisfied,
subject to minimality, meaning that we search for $Y(u)$ which ensures
the axioms are fulfilled with the minimum number of poles and zeros on the
physical strip (the region $0\leq{\rm Re}(u)\leq1$).

The unitarity constraint which follows from the hypothesis that the space
of states of the theory is complete can be satisfied by virtue of the
completeness relation \CR\ if $f(u)=cu$, for some constant $c$, and
$$
Y(u)Y(-u)=1/\varrho\left(f(u)\right).
\nfr{UNIT}
Crossing symmetry of the $S$-matrix relates the process \KP\ with the
the process $K_{cd}+{\overline K}_{db}\rightarrow{\overline K}_{ca}+
K_{ab}$, where ${\overline K}$ is the charge conjugate kink. Notice, that
for the $b_m$, $c_m$ and $d_m$ algebras if $a-b\in\Sigma$ then
$b-a\in\Sigma$,
due to the fact that the representations are real, therefore we may take
${\overline K}_{ab}\equiv K_{ab}$. In these cases crossing symmetry requires
$$
{\widetilde S}\left(\left.\matrix{a&b\cr c&d\cr}\right\vert u\right)=
{\widetilde S}\left(\left.\matrix{c&a\cr d&b\cr}\right\vert 1-u\right),
\efr
which may be satisfied by virtue of \CS\ if $f(u)=u$ and
$$
Y(1-u)=Y(u).
\nfr{CROSS}

The situation for $a_{m-1}$ is somewhat different since the vector
representation
is not conjugate to itself. However, this case has been dealt with
elsewhere [\Ref{MSQ},\Ref{VF}]. One finds that a crossing symmetry relation
can be
satisfied when kinks in the conjugate vector representation are included in
the spectrum and the minimal solution for $Y(u)$ is
$$\eqalign{
Y_{m,k}&(u)=\exp\left\{
\int_0^\infty{dx\over x}{2\,{\rm sinh}(mux/2)\over{\rm sinh}[(k+m)x]{\rm sinh}
(mx)}\right.\cr
&\times\left({\rm cosh}(kx){\rm cosh}(mxu/2)-{\rm cosh}[(m+k-2)x]
{\rm cosh}[mx(u/2-1)]\right)\Big\},\cr}
\nfr{YFA}
where $k$ is a parameter related to $\omega$ as in \RESM.
$Y_{m,k}(u)$ can also be expressed in terms of products of Gamma
functions [\Ref{VF}]. The function $Y_{m,k}(u)$ satisfies two important
identities:
$$
Y_{m,k}(u)Y_{m,k}(-u)={\sin^2\omega\over\sin(\omega+\lambda u)
\sin(\omega-\lambda u)},
\nfr{RELO}
and
$$
Y_{m,k}(1-u)Y_{m,k}(1+u)={\sin^2\omega\over\sin(\lambda-\lambda u)
\sin(\lambda+\lambda u)},
\nfr{RELT}
where as before $\omega=\pi/(m+k)$ and $\lambda=m\omega/2$.

Using these two relations we can now write down $S$-matrices for the other
algebras by noting that a solution of \UNIT\ and \CROSS\ is
$$
Y(u)=Y_{tg,tk}(u)Y_{tg,tk}(1-u){\sin\lambda\over\sin\omega},
\quad{\rm for}\ b_m,c_m,d_m.
\nfr{YFO}
Summarizing we have
$$\eqalign{
{\widetilde S}\left(\left.\matrix{a&b\cr c&d\cr}\right\vert u\right)&=
W\left(\left.\matrix{a&b\cr c&d\cr}\right\vert u\right)\left({
G_aG_d\over G_bG_c}\right)^{u/2}\cr
&\times\cases{Y_{m,k}(u),&$
\quad {\rm for}\ a_{m-1},$\cr
Y_{tg,tk}(u)Y_{tg,tk}(1-u){\sin\lambda\over\sin\omega},\quad&$
{\rm for}\ b_m,c_m,d_m.$\cr}}
\nfr{AN}

It is straightforward to show using \IC\ and the
fact that $Y_{tg,tk}(0)=1$ and $Y_{tg,tk}(1)=\sin\omega/\sin\lambda$ that
$$
{\widetilde S}\left(\left.\matrix{a&b\cr c&d\cr}\right\vert0\right)=
\delta_{bc}.
\nfr{ZM}

Next we turn to an investigation of the analytic structure structure of the
above $S$-matrices. The factor $Y_{tg,tk}(u)$ has
no poles or zeros on the physical strip
which means that there are no bound state resonances.
So the $S$-matrix defines a complete theory for the cases $b_m$, $c_m$ and
$d_m$ with kinks associated to the vector representation of the algebra.
For the case $a_{m-1}$ the $S$-matrix does not make sense on its own because
the requirement of crossing symmetry means that kinks in the conjugate
representation should be included in the spectrum.

We now consider the case of the $a_{m-1}$ theory in more detail. In order
to satisfy crossing symmetry the conjugate vector representation
must be included in the spectrum. This is achieved in a dynamical way
by a generalization of the original Ansatz \AN:
$$
S\left(\left.\matrix{a&b\cr c&d\cr}\right\vert u\right)=
X(u)
{\widetilde S}\left(\left.\matrix{a&b\cr c&d\cr}\right\vert u\right),
\nfr{TRY}
where the additional CDD factor $X(u)$ does not upset the unitarity
or crossing symmetry relations but provides an additional pole
on the physical strip. In order to motivate the form of the scalar factor
consider the spectral decomposition of the $R$-matrix (see appendix A):
$$
\rc(u)={\sin(\omega-\lambda u)\over\sin\omega}{\Bbb P}_{2\omega_1}+{\sin(
\omega+\lambda u)\over\sin\omega}{\Bbb P}_{\omega_2},
\nfr{APR}
where ${\Bbb P}_{2\omega_1}$ and ${\Bbb P}_{\omega_2}$ are the quantum group
invariant projectors onto the representations with highest weights
$2\omega_1$ and $\omega_2$ which appear in the tensor product of two vector
representations. The idea is that $X(u)$ should have a simple pole at
$u=\omega/\lambda=2/m$  at the place where
the $R$-matrix projects onto the
second fundamental representation: $\rc(u)\propto{\Bbb P}_{\omega_2}$.
This determines the form of the scalar factor:
$$
X(u)={\sin\left({\pi u\over2}+{\pi\over m}\right)\over\sin\left(
{\pi u\over2}-{\pi\over m}\right)}.
\nfr{ACDD}
As a consequence a particle transforming in the second fundamental
representation in included in the spectrum and the
$S$-matrix elements of this new state can then be found using
the bootstrap equations. The process is then repeated: the new $S$-matrix has
simple poles on the physical strip
which are interpreted in terms of states propagating in the direct or
crossed channels. For the unrestricted models
it was shown in [\Ref{MSQ}] that the procedure
terminates on a finite set of particles corresponding
to all the fundamental representations of $a_{m-1}$ (in particular the
conjugate vector representation) if the coupling constant
satisfies the inequality $\omega<2\pi/m$. Notice that the restricted models,
for which $\omega=\pi/(m+k)$, all lie in this region. The most
important comment to make about this procedure is that the positions of the
poles are dictated by the fusion structure of the solutions of the YBE, which
in turn determines the ratios of the masses to be
$$
m_a\propto\sin\left({\pi a\over m}\right),\quad a=1,2,\ldots,m-1.
\nfr{AMAS}
Here the particle with mass $m_a$ transforms in the $a^{\rm th}$ fundamental
representation. This is the mass spectrum of the $S$-matrix of the $a_{m-1}$
affine Toda field theory. This is not surprising because the extra CDD factor
in \TRY\ is the minimal\note{By minimal we mean the part of the
Toda $S$-matrix which is independent of the coupling constant.}  $S$-matrix
of the aforementioned theory for the particle associated to
the vector representation. In fact the poles on the physical strip of the
non-diagonal theory are completely determined by the minimal Toda $S$-matrix
since the rest of the $S$-matrix has no poles or zeros
on the physical strip. It is
important to point out that the minimal Toda scalar $S$-matrix has double
poles on the physical strip, but these are
explained in terms of the anomalous thresholds produced by `on-shell' diagrams
[\Ref{TOD}]. We will discuss the full solution of the bootstrap equations in
the next section.

We can now try to construct $S$-matrices for the other classical Lie
algebras along the same lines as for the $a_{m-1}$ algebra.
The first thing to consider is the spectral decomposition of the
solution of the Yang-Baxter equation for vector-vector scattering. It is
known that [\Ref{JIM}]
$$\eqalign{
\rc(u)=&
{\sin(\lambda-\lambda u)\sin(\omega-\lambda u)\over
\sin\omega\sin\lambda}{\Bbb
P}_{2\omega_1}+{\sin(\lambda-\lambda u)\sin(\omega+\lambda u)\over
\sin\omega\sin\lambda}{\Bbb P}_{
\omega_2}\cr
&\qquad\qquad\qquad+
{\sin(\lambda+\lambda u)\sin(\omega+\sigma\lambda u)\over\sin\omega\sin
\lambda}{\Bbb P}_0,\cr}
\nfr{OTHP}
where $\sigma$ is $=-1$ for $c_m$ and $1$ otherwise.
So if we wish to include the second fundamental representation in the
spectrum we would require a simple pole at $u=\omega/\lambda=2/tg$; however,
notice that for $c_m$ the residue at this point projects onto the
irreducible representation, whereas for the other algebras the residue
projects onto a reducible representation consisting of the second
fundamental representation plus a trivial representation: we are
forced to conclude that this second particle transforms in this reducible
representation in these cases. A simple pole at this point can be introduced by
including a CDD factor
$$
X(u)={\sin\left({\pi u\over2}+{\pi\over tg}\right)\sin\left(
{\pi\over tg}
+{\pi\over2}-{\pi u\over2}\right)\over
\sin\left({\pi u\over2}-{\pi\over tg}\right)\sin\left(
{\pi\over tg}-{\pi\over2}+{\pi u\over2}\right)}
\nfr{CDDF}
which also has a simple pole corresponding to the process where the second
particle is exchanged in the crossed-channel as demanded by crossing
symmetry. The form of $X(u)$ ensures
that it satisfies unitarity and crossing independently of the non-diagonal
part of the $S$-matrix.

So the proposal for the $S$-matrix for vector-vector scattering has the form
$$
S(u)=X(u){\widetilde S}(u),
\efr
where ${\widetilde S}(u)$ is the $S$-matrix in \AN\ and $X(u)$ is the CDD
factor \ACDD\ or \CDDF\ which is the minimal $S$-matrix for the vector-vector
scattering
of an associated affine Toda theory. The associated Toda theories are the
following:
$$
a_{m-1}\rightarrow a_{m-1}^{(1)},\ \
b_m\rightarrow a_{2m-1}^{(2)},\ \ c_m\rightarrow d_{m+1}^{(2)},\ \
d_m\rightarrow d_m^{(1)}.
\nfr{RELTODA}
The associated algebra is in fact the one whose Cartan matrix is
equal to the transpose of the Cartan matrix of the
untwisted affinization of the original
algebra. We emphasize that $X(u)$ is not the Toda $S$-matrix itself but rather
the part which is independent of the coupling constant (for the non-simply
laced algebras we refer to the na\"\i ve $S$-matrix elements that one would
write down for the particles with the classical mass ratios [\Ref{TOD}] as
opposed to the actual $S$-matrix of the Toda theory [\Ref{NLS}]).

On the basis of this there are serious objections to the $S$-matrices that we
have proposed. Firstly, for $b_m$ and $d_m$ the vector representation
cannot be considered as the elementary particle from which all the others
follow as bound states because the spinor representations could not
produced in this way. One might think that one could define a consistent
$S$-matrix which does not include the spinor representations and for
which the vector representation is elementary, however
this cannot be so because the spinor makes an appearance in the
correct interpretation of the higher order poles. In other
words such an $S$-matrix would have some unexplained higher order poles
and must consequently be rejected.\note{It is just
possible for the matrix
factor of the $S$-matrix to provide some judicious zeros, but
this is not the case.} The second thought is that if
one starts with the spinor representations in a similar way
then the $S$-matrices that we have written down would follow by fusion, in
other words we have only written down part of the $S$-matrix.

There is, however, a
more serious objection which applies to the non-simply-laced theories.
The objection arises from the fact that for the non-simply-laced
algebras the minimal Toda $S$-matrix part of the conjecture does not
by itself define a consistent $S$-matrix. The reason is that as written
down the scalar $S$-matrices for the $a^{(2)}_{2m-1}$ and $d^{(2)}_{m+1}$
Toda theories have multiple
poles on the physical strip which cannot be explained by the spectrum
of particles which follows from the classical Toda Lagrangian [\Ref{TOD}]:
$$\eqalign{
a^{(2)}_{2m-1}&:\ m_j\propto\cases{1 &$j=m$\cr 2\sin\left[{j\pi\over2m-1}
\right]\ \ &$j=1,2,\ldots,m-1,$\cr}\cr
d^{(2)}_{m+1}&:\ m_j\propto\sin\left[{
j\pi\over2(m+1)}\right],\quad j=1,2,\ldots,m.\cr}
\nfr{MSS}
The best way to see this is to consider the $a^{(2)}_{2m-1}$ and
$d^{(2)}_{m+1}$ $S$-matrices as a subset of the minimal $S$-matrix of
a larger Toda theory. The $a^{(2)}_{2m-1}$ $S$-matrix is
the $d^{(1)}_{2m}$ $S$-matrix for the subset of particles $\{2,4,\ldots,2m\}$,
whilst the $d^{(2)}_{m+1}$ $S$-matrix is the $d^{(1)}_{m+2}$ for the subset
of particles $\{1,2,\ldots,m\}$ (i.e. without the spinor and anti-spinor
particles). The point now is that the sub-$S$-matrix has multiple poles on
the physical strip which require the full spectrum of the larger theory in
order to be explained.

In fact the resolution of this problem for the non-simply-laced Toda
theories is surprisingly subtle. What happens is that the ratios of the
particle masses get renormalized in a non-trivial way such that the mass
ratios depend on the coupling constant with the consequence that the analytic
structure becomes modified. The new $S$-matrix no longer suffers from the
pathologies of the na\"\i ve one. The non-diagonal $S$-matrices
under discussion would, on the face of it,
suffer the same fate as their na\"\i ve non-simply-laced
Toda counterparts, since it is the na\"\i ve Toda factor that provides
the poles on
the physical strip; however, in this case we cannot hypothesize
that the particle masses get renormalized in some non-trivial way
because the positions of the poles must match the fusing of the
${\widetilde S}$ factor and this fusing structure is completely rigid: the
positions of the poles are determined completely.

If the $S$-matrices conjectured in \TRY\ are to be completely consistent then
the part arising
from the fusion of ${\widetilde S}$ must provide some appropriate
zeros in order to cancel the unwanted poles of the Toda theory factor.
In the next section we show how this can occur by solving the bootstrap
equation for the $a_{m-1}$ and $c_m$ cases. The obstruction for dealing with
the $b_m$ and $d_m$ algebras in the same way, apart from the problem that
the vector particle is not elementary, is that the relevant spectral
decompositions of the $R$-matrices are not known.

$S$-matrices associated to the vector representations of the classical Lie
algebras have recently been proposed in [\Ref{GEP}]. However, this reference
uses an incorrect form for the spectral decomposition \OTHP\ and misses the
crucial factor of $t$ in the above equations which render the result invalid
for $c_m$. It also asserts that the masses of the theories are those of the
corresponding classical Toda theory; whereas our results show that they are
actually those of the related classical Toda theories \MSS.

\chapter{The solution of the bootstrap for $a_{m-1}$ and $c_m$}

In this section we will explicitly solve the bootstrap for the $a_{m-1}$ and
$c_m$ theories. This involves the resolution of two interlocking problems.
Firstly we must find the spectral decompositions of the $R$-matrix on the
fundamental representations and then we must show how the scalar factor
provides the correct analytic structure on the physical strip.

Fortunately, for the representations that are necessary to construct the
$S$-matrices, there is well-established technology for finding the spectral
decompositions of the associated $R$-matrices. We explain how the spectral
decompositions are found in appendix A. However, knowing the spectral
decompositions is not enough for our purposes, they only tell us the form of
the $R$-matrix up to a scalar function of the spectral parameter. In fact,
we must solve the bootstrap equations for which the overall scalar factors
are determined and indeed crucial for the correct form of the $S$-matrix. We
derive these scalar factors in appendix B.

Our results are the following. We denote by $\rc^{ab}(u)$ the $R$-matrix
between the $a^{\rm th}$ and $b^{\rm th}$ fundamental representations
(where the highest weight of the $a^{\rm th}$ fundamental representation is
$\omega_a=e_1+e_2+\cdots+e_a$), and we
choose without loss of generality $b\geq a$. For $a_{m-1}$ we have
$$
\rc^{ab}(u)=Z^{ab}_1(u)\sum_{k=0}^{{\rm min}(m-b,a)}(-)^{k+1}
\rho^{ab}_k(u){\Bbb P}_{\omega_{b+k}+\omega_{a-k}},
\nfr{ARSD}
with
$$\eqalign{
\rho^{ab}_k(u)&=\prod_{p=1}^{k}\{2p+b-a\}\prod_{p=k+1}^{{\rm min}
(m-b,a)}\{-2p-b+a\},
\cr
Z^{ab}_1(u)&=\prod_{j=1}^a\prod_{k=1}^{b-1}\{2j+2k-a-b\}
\prod_{p={\rm min}(m-b,a)+1}^{a}\{-2p-b+a\}.\cr}
\efr
where we take $\omega_m=\omega_0=0$ and
$$
\{x\}={\sin\left(\omega x/2+\lambda u\right)
\over\sin\omega}.
\nfr{BR}
For $c_m$ we have
$$
\rc^{ab}(u)=Z^{ab}_2(u)\sum_{j=0}^{{\rm min}(m-b,a)}
\sum_{k=0}^{a-j}(-)^{j+k}
\rho^{ab}_{jk}(u){\Bbb P}_{\omega_{b+j-k}+\omega_{a-j-k}},
\nfr{OTRSD}
where
$$\eqalign{
\rho^{ab}_{jk}(u)=&
\prod_{p=1}^j\{2p+b-a\}\prod_{q=1}^k\{2(m+1)+2q-a-b\}\cr
&\quad\times\prod_{p=j+1}^{{\rm min}(m-b,a)}
\{-2p-b+a\}\prod_{q=k+1}^a\{-2(m+1)-2q+a+b\}.\cr
Z^{ab}_2(u)=&{\left(\sin\lambda\over\sin\omega\right)}^{ab}
\prod_{j=1}^a\prod_{k=1}^{b-1}\{2j+2k-a-b\}\{a+b-2(m+1)-2j-2k\}
\cr &\qquad\times\prod_{p={\rm min}(m-b,a)+1}^a\{-2p-b+a\}.\cr}
\efr

The full $S$-matrix is then equal to
$$
{\widetilde S}^{ab}_{(k)}(u)=Y^{ab}(u)\rc^{ab}(u),
\nfr{EFF}
with
$$
Y^{ab}(u)=\prod_{j=1}^{a}
\prod_{k=1}^{b}Y\left(u+{1\over tg}(2j+2k-a-b-2)\right).
\nfr{YD}
where $Y(u)$ is given by \YFA\ for $a_{m-1}$ and \YFO\ for $c_m$.
In \EFF\ we have explicitly indicated the dependence on the parameter $k$
via $\omega$ in \RESM.
The normalization factors $Z_1^{ab}(u)$ and $Z_2^{ab}(u)$ arise from solving
the bootstrap equations starting with $\rc(u)$ and are crucial for the correct
form of the $S$-matrix since they can provide zeros on the physical strip and
ensure that the unitarity condition is satisfied.
To illustrate the latter point we notice that
$$
\rc^{ab}(u)\rc^{ba}(-u)=\prod_{j=1}^{a}
\prod_{k=1}^{b}\varrho\left(u+{1\over tg}(2j+2k-a-b-2)\right),
\efr
where $\varrho(u)$ is defined in \RHOD. So using \YD\ along with \RELO\
and \RELT\ we deduce that
$$
{\widetilde S}^{ab}(u){\widetilde S}^{ba}(-u)=I^b\ot I^a.
\efr

We are now in a position to investigate the analytic structure of
${\widetilde S}^{ab}_{(k)}(u)$ on the physical strip. For $a_{m-1}$ there
are no poles or zeros on the physical strip, whereas
for $c_m$ there are no poles but if $a+b>m$ there is a series
of simple zeros at
$$
u={a+b-2j+2\over2(m+1)},\quad j=1,2,\ldots,a+b-m.
\nfr{ZEROS}

A consistent $S$-matrix is made by appending a suitable minimal Toda
factor which provides the necessary pole structure.
Generalizing the case for $a_{m-1}$ one is tempted to try the
Gross-Neveu (GN) type Ansatz
$$
S^{ab}_{(k)}(u)=X^{ab}(u){\widetilde S}^{ab}_{(k)}(u),
\nfr{GGN}
where $X^{ab}(u)$ is the minimal Toda factor of the associated algebra
\RELTODA. However, we shall
argue below that this is inconsistent for the non-simply-laced algebras
due to the appearance of spurious poles on the physical strip which
cannot be explained. Rather we shall find that a consistent $S$-matrix
is given by the tensor product form which generalizes the $S$-matrix
of the Principal Chiral Model (PCM):
$$
S^{ab}_{(k,l)}(u)=X^{ab}(u){\widetilde S}^{ab}_{(k)}(u)\otimes
{\widetilde S}^{ab}_{(l)}(u).
\nfr{GPCM}
For this $S$-matrix the particles transform in the reducible
representations $V_{\omega_a}
\otimes V_{\omega_a}$ where $V_{\omega_a}$ is the
representation of the algebra with highest weight $\omega_a$.

Let us verify the above statements for algebras $a_{m-1}$ and $c_m$.
The former case is easy to discuss. The minimal Toda factor is
associated to the algebra $a^{(1)}_{m-1}$:
$$
X^{ab}(u)=\prod_{j=|a-b|+1\atop{\rm step}\ 2}^{a+b-1}(j+1)(j-1),
\efr
with the notation
$$
(j)={\sin\left({\pi u\over2}+{\pi j\over2tg}\right)\over\sin\left(
{\pi u\over2}-{\pi j\over2tg}\right)},
\nfr{BRANOT}
where $tg=m$ in this case.
The pole in $X^{ab}(u)$ at $(a+b)/m$ (if $a+b<m$) or
$2-(a+b)/m$ (if $a+b>m$) corresponds to particle $a+b$ or $a+b-m$,
respectively, in the direct channel. One can verify directly using \ARSD\ that
the residues at these poles are ${\Bbb P}_{\omega_{a+b}}$ and ${\Bbb
P}_{\omega_{a+b-m}}$, respectively, as required for consistency.
The pole at $|a-b|/m$ corresponds
to the particle $|a-b|$ in the crossed channel. The element
$S^{ab}(u)$ also exhibits double poles at
$$
u={a+b-2k\over m},\quad k=1,2,\ldots,{\rm min}(a,b)-1.
\nfr{DPA}
These double poles are understood in terms of the Coleman-Thun mechanism
[\Ref{OSD}] which was originally posited in order to explain the double
poles in the sine-Gordon $S$-matrix. Essentially, the poles are
Landau singularities of the ordinary sort, which appear as poles in
two-dimensions and branch-points in four-dimensions, and are associated to
`on-shell' diagrams. In the present example,
figure 1 gives the on-shell diagram causing the $k^{\rm th}$ double pole in
\DPA\ for the case $a,b\leq m/2$ (other cases are found by crossing).
We therefore conclude that both the generalized
GN- and PCM-type $S$-matrices are consistent for $a_{m-1}$.

\sjump
\midinsert
\centerline{
\psfig{figure=ac.eps,width=2.5in}}
\bjump
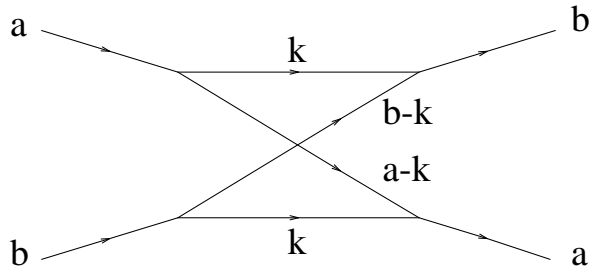
\centerline{1. On-shell diagram giving double pole.}
\endinsert

The scalar factor for $c_m$ is the $d^{(2)}_{m+1}$ minimal Toda
$S$-matrix. This $S$-matrix can be understood as the $S$-matrix of the
$d_{m+2}^{(1)}$ excluding the spinor and anti-spinor particles. So
$$
X^{ab}(u)=\prod_{j=|a-b|+1\atop{\rm step}\ 2}^{a+b-1}(j+1)(j-1)(2g-j+1)
(2g-j-1),
\efr
where $g=m+1$ is the dual Coxeter number of $c_m$ and the bracket notation
is defined in \BRANOT. So $X^{ab}(u)$ in
general exhibits poles of order 1,2,3 and 4 on the physical strip.
Consider a GN-type $S$-matrix. The analytic structure of $S^{ab}_{(k)}(u)$
is deduced by combining that of $X^{ab}(u)$ with the zeros \ZEROS\ of
the ${\widetilde S}^{ab}(u)$ part.
Rather than analyzing the $S$-matrix in full detail we shall exhibit a
singularity which has no explanation in terms of the particle spectrum.
To this end one notices that $X^{a,m+1-a}(u)$ has a double pole at $u=1/2$
which is explained in terms of the on-shell diagrams in figure 2
involving the spinor
and anti-spinor particles of the $d_{m+2}^{(1)}$ theory.\note{I am grateful
to Patrick Dorey for explaining this point.}
${\widetilde S}^{ab}(u)$ has a simple zero at $u=1/2$; hence
$S^{ab}_{(k)}(u)$ has a simple pole at $u=1/2$ which cannot be explained in
terms of the particle spectrum.
At this stage we are forced to conclude that there cannot be a consistent
$S$-matrix of the GN-type.\note{It might be possible to enlarge
the set of particles and then close the bootstrap, however, the new particle
would transform in a reducible representation involving non-fundamental
representations.}

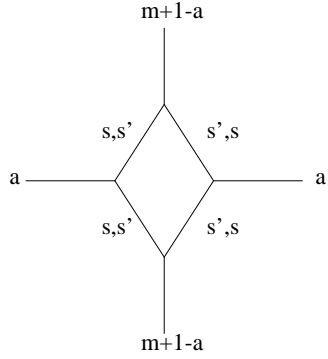
\sjump
\midinsert
\centerline{
\psfig{figure=spin.eps,height=1.5in}}
\bjump
\centerline{2. On-shell diagram giving double pole in $X^{a,m+1-a}(u)$
for $d_{m+2}^{(1)}$.}
\endinsert

The PCM-type of $S$-matrix, however, does not have this spurious pole since
there is an extra zero to cancel it. The analytic structure on the physical
strip is illustrated in figure 3 where the crosses represent poles.

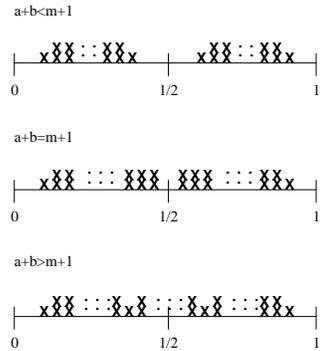
\sjump
\midinsert
\centerline{
\psfig{figure=sing.eps,height=1.5in}}
\bjump
\centerline{3. Poles in the PCM-type $S$-matrix for $c_m$.}
\endinsert

Notice that there are only simple poles and
double poles. Some of the
simple poles can be understood in terms of direct
or crossed channel resonances of the particles and we  consider these ones
first. It is useful to distinguish three cases.

(i) $a+b<m+1$. In this case there are four simple poles. The simple poles at
$(a+b)/2(m+1)$ and $1-|a-b|/2(m+1)$ correspond to the exchange of particles
$a+b$ and $|a-b|$ in the direct channel,
respectively, with residues proportional to ${\Bbb P}_{\omega_{a+b}}$
and ${\Bbb P}_{\omega_{|a-b|}}$.
The simple poles at $|a-b|/2(m+1)$ and $1-(a+b)/2(m+1)$ correspond to the
exchange of $|a-b|$ and $a+b$ in the
crossed channel, respectively.

(ii) $a+b=m+1$. In this case there are two simple poles. The simple
poles at $|a-b|/2(m+1)$ and $1-|a-b|/2(m+1)$ correspond to the exchange of
particle $|a-b|$ in the crossed and direct channel, respectively. In the latter
case the residue is proportional to ${\Bbb P}_{\omega_{|a-b|}}$.

(iii) $a+b>m+1$. In this case there are four simple poles, however,
only two of them correspond to direct or cross channel resonances. The simple
poles at $|a-b|/2(m+1)$ and $1-|a-b|/2(m+1)$ correspond to the exchange of
particle $|a-b|$ in the crossed and direct channel, respectively. In the latter
case the residue is proportional to ${\Bbb P}_{\omega_{|a-b|}}$.

We now turn to the remaining singularities. Firstly if $a+b>m+1$ there are
simple poles at $(a+b)/2(m+1)$ and $1-(a+b)/2(m+1)$. Secondly
there are double poles are in the union of the set
$$
u={a+b-2k\over 2(m+1)},\quad k=1,2,\ldots,{\rm min}(a,b)-1,
\efr
and its crossed version, with the exceptions that for $a+b>m+1$
$u=(a+b)/2(m+1)$ and $u=1-(a+b)/2(m+1)$ are simple rather than double
poles and when $a+b=m+1$ $u=1/2$ is regular.

There exists at least one on-shell diagram at the positions of the double
poles, which involve particles in the spectrum and which lead to
double poles on kinematical grounds. These on-shell diagrams
are the same as those in figure 1 (plus the crossed versions).
If $a+b>m+1$ one would expect to see triple poles at $(a+b)/2(m+1)$
and $1-(a+b)/2(m+1)$ coming from the factor $X^{ab}(u)$, whereas in
fact the $S$-matrix only has simple poles.
It is rather unconventional to have an $S$-matrix, some of whose
simple poles do not correspond to direct or cross channel resonances.
However, [\Ref{CDS}] discusses a recent example of such an eventuality in
non-simply-laced Toda theories. These strange occurrences
are explained by a generalization of the Coleman-Thun mechanism
[\Ref{OSD}], in the sense that there are on-shell diagrams
which at first sight seem to lead to higher order poles but which on closer
inspection actually only lead to simple poles because some sub-$S$-matrix
element in the diagram has a zero. We shall find that a similar mechanism
is at work in the example under discussion, except that here
the softening of the singularity is not due to some sub-$S$-matrix
element having a zero but rather is intimately bound up with the fact that
the particles carry internal quantum numbers and one must sum all the
quantum numbers of particles on internal lines.

The diagram which looks like it would lead to a triple pole
at $(a+b)/2(m+1)$ (and $1-(a+b)/2(m+1)$ by crossing) is illustrated in
figure 4 (with the incoming particles coming from the left),
where the grey blob is the $S$-matrix element $S^{m+1-a-b,
m+1-a-b}(u)$ evaluated at $u\equiv u_0=1-(a+b)/2(m+1)$ and the
black blobs represent projection operators.

\sjump
\midinsert
\centerline{
\psfig{figure=tri.eps,height=1.5in}}
\bjump
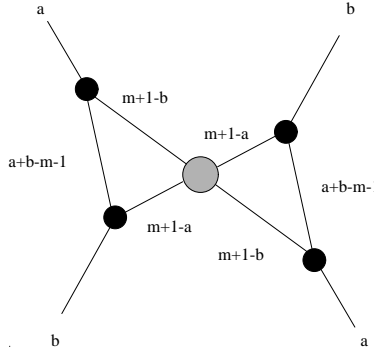
\centerline{4. On-shell diagram for $a+b>m+1$.}
\endinsert

As it stands, standard kinematical
arguments would indicate that the diagram
yields a double pole. However the $S$-matrix element $S^{m+1-a-b,
m+1-a-b}(u)$ has a simple pole at $u_0$ corresponding to the
exchange of $2(m+1)-a-b$ in the direct channel; hence the expected
singularity is a triple pole. However, this argument is too na\"\i ve
and we must investigate the residue more closely.
The behaviour of the internal $S$-matrix
element in the vicinity of the pole to $O(u-u_0)$ is
$$\eqalign{
&S^{m+1-a-b,m+1-a-b}(u)=\cr
&\qquad{1\over u-u_0}\left(a{\Bbb
P}'+(u-u_0)
\sum_{\mu}b_\mu{\Bbb P}_\mu\right)\bigotimes\left(a{\Bbb
P}'+(u-u_0)\sum_\mu b_\mu{\Bbb P}_\mu\right),\cr}
\efr
where ${\Bbb P}'={\Bbb P}_{\omega_{2(m+1)-a-b}}$,
$a$ and $b_\mu$ are constants and the sums are over the
highest weights of the representations that appear in the tensor
product $V_{\omega_{m+1-a-b}}\otimes V_{\omega_{m+1-a-b}}$.
The relevant observation is that the
representation with highest weight $\omega_{2(m+1)-a-b}$
does not actually appear in the
tensor product $V_{\omega_a}\otimes V_{\omega_b}$ (remember that
$a+b>m+1$). So if we isolate the group-theoretic contributions
from one of the factors in the tensor product to the
$S$-matrix we conclude that the terms of the form
${\Bbb P}_0{\Bbb P}'{\Bbb P}_0$,
where ${\Bbb P}_0$ represents the product of the two projection operators
on the external legs and hence is a homomorphism from $V_{\omega_a}\otimes
V_{\omega_b}$ to $V_{\omega_{m+1-a-b}}\otimes V_{\omega_{m+1-a-b}}$,
must vanish.
Hence when the diagram is evaluated one actually picks up the contribution
from the central $S$-matrix element at $O(u-u_0)$, rather than at
$O(u-u_0)^{-1}$. In other words the contributions from the more
singular diagrams are zero
after one has summed over all the intermediate states. So the actual
singularity is a simple pole rather than a triple pole since each triangle
contributes a simple pole from its kinematical factors.

It is important to be aware of the fact that the
$S$-matrix $X^{ab}(u)$ also has fourth order poles which can be explained
solely in terms of the set of particles in the $d_{m+1}^{(2)}$ spectrum.
These singularities are reduced to poles of second order in the PCM-type
$S$-matrix. In order for this to occur the residues of the diagrams
corresponding to the fourth order poles must vanish. Such an eventuality
is possible because the particles carry internal quantum numbers ---
as we have seen from the preceding arguments. Unfortunately, an analysis of
the residues is a rather formidable problem
and is not within the scope of the present article.

Before we move on to discuss the rational limits of the $S$-matrices
we first pause to consider the $d_m$ and $b_m$ cases.
As we have pointed out for these algebras the vector is
not the elementary particle, rather it is the spinor (and anti-spinor)
particle. Nevertheless our $S$-matrices should generate a subset of the full
$S$-matrix. One finds that in these cases the bound states are not
associated to the fundamental representations, as in the $a_{m-1}$ and
$c_m$ cases. In fact the $a^{\rm th}$ particle transforms in the
reducible representation [\Ref{OW}]
$$
W_a=\bigoplus_{j=0}^{a-2j\geq0}V_{\omega_{a-2j}}.
\nfr{REP}
Unfortunately the spectral decompositions have not
been found for these representations.

\chapter{The rational limit}

The rational limit of the $S$-matrices are obtained by taking $k\rightarrow
\infty$ and the resulting $S$-matrices
are those of some well-known quantum field theories.
Furthermore one can show by explicit computation that
the `important'
analytic structure of the $S$-matrix is not affected by the limit, in the
sense that no poles or zeros from the ${\widetilde S}$ factor can wander
onto or off the physical strip. The rational $S$-matrices are actually
invariant under the group associated to the Lie algebra in question
(on the contrary the trigonometric $S$-matrices of the last section are
invariant under the quantum group).

The rational limit of the $R$-matrices is easily obtained by taking the
$k\rightarrow\infty$ limits of \ARSD\ and \OTRSD, which means replacing
\BR\ with
$$
\{x\}=(x+tgu)/2,
\nfr{NBR}
and taking the ${\Bbb P}$'s to be the Lie algebra, rather than quantum group,
invariant homomorphisms. This latter limit is taken because as
$k\rightarrow\infty$ the deformation parameter of the quantum group
$q\rightarrow-1$ and so the quantum group reduces to the Lie algebra.

The factor $Y_{tg,tk}(u)$ also has a good limit:
$$
Y_{tg,\infty}(u)={1\over tg}{\Gamma\left(1-{u\over2}\right)\Gamma\left(
{1\over tg}+{u\over2}\right)\over\Gamma\left(1+{u\over2}\right)
\Gamma\left(1+{1\over tg}-{u\over2}\right)}.
\nfr{YINF}
Using these results we find for vector-vector scattering
$$
{\widetilde S}_{(\infty)}(u)=
{\Gamma\left(1-{u\over2}\right)\Gamma\left({1\over m}+{u\over2}\right)
\over\Gamma\left(1+{u\over2}\right)\Gamma\left({1\over m}-{u\over2}\right)}
\left[{\Bbb P}_{2\omega_1}+\left({{2\over m}+u\over{2\over m}-u}\right)
{\Bbb P}_{\omega_2}\right],
\nfr{RAN}
for $a_{m-1}$ and
$$\eqalign{
{\widetilde S}_{(\infty)}(u)=
&{\Gamma\left(1-{u\over2}\right)\Gamma\left({1\over tg}+{u\over2}\right)
\Gamma\left({1\over2}+{u\over2}\right)\Gamma\left({1\over tg}+{1\over2}-
{u\over2}\right)\over
\Gamma\left(1+{u\over2}\right)\Gamma\left({1\over tg}-{u\over2}\right)
\Gamma\left({1\over2}-{u\over2}\right)\Gamma\left({1\over tg}+{1\over2}
+{u\over2}\right)}\cr
&\quad\times\left[{\Bbb P}_{2\omega_1}+\left({{2\over tg}+u\over{2\over tg}
-u}\right){\Bbb P}_{\omega_2}+\left({1+u\over1-u}\right)
\left({{2\over tg}+\sigma u\over{2\over tg}
-u}\right){\Bbb P}_0\right],\cr}
\nfr{ROT}
for the other algebras.

A careful comparison of the rational limits
of our $S$-matrices \RAN\ and \ROT\ with those of the principal chiral
model [\Ref{BOOT}],\note{In comparing our expressions with this reference
it is important to notice that the $S$-matrices defined there are equal to
ours up to a permutation of the outgoing particles.} in
which the particles transform in a tensor product of fundamental
representations of the algebra, confirms that
$$
S_{\rm PCM}^{ab}(u)=S^{ab}_{(\infty,\infty)}(u)=
{\widetilde S}_{(\infty)}^{ab}(u)\otimes{\widetilde S}_{(\infty)}^{ab}(u)X^{ab}
(u),
\efr
for $a_{m-1}$ and $c_m$ (and $b_m$ and $d_m$ for $a=b=1$). This relation
explains our use the nomenclature `PCM-type' for the $S$-matrix of \GPCM.

For $a_{m-1}$ we can also consider the $S$-matrix
$$
S_{GN}^{ab}(u)=S^{ab}_{(\infty)}(u)={\widetilde S}^{ab}_{(\infty)}(u)X^
{ab}(u),
\efr
which is the $S$-matrix of the $SU(m)$ Gross-Neveu model [\Ref{BOOT},\Ref{GN}].

The fact that the Gross-Neveu $S$-matrix for the non-simply-laced algebras
violates the bootstrap was noted in [\Ref{BOOT}]. It has been claimed
[\Ref{BOOT}] that the $S$-matrix of the principal chiral model for $c_m$
violates the bootstrap due
to the appearance of simple poles with no explanation in terms of
direct or cross channel poles. However, the analysis of the last
section shows how this problem is resolved
and the poles can be properly understood via a generalization of the
Coleman-Thun mechanism.

\chapter{The situation at $k=1$}

In this section we consider what happens to the $S$-matrices when the
level $k=1$. It is known from the study of quantum groups that when
$q$ is a root of unity, so $k$ is an integer, that the representation
theory has to modified. This modification is implemented automatically
by moving to the IRF picture and means that the usual rule for decomposing
tensor products is truncated on the set of representations which correspond
to highest weights of level $k$. This would entail appropriate modifications
of the spectral decompositions. When $k=1$ only level one representations
survive. It is significant that in the cases where the particles are
associated to higher level representations, i.e. for the $b_m$ and $d_m$
theories where there are representations at level 2, then the
representations \REP\
are reducible and contain a level 1 component, so that all the particles remain
in the spectrum.\note{This observation seems
also to be true for the known solutions of the YBE for the exceptional
algebras as well.}

As we mentioned, the appropriate way to write down the $S$-matrix for
$k=1$ is via the IRF formalism. At $k=1$ the allowed weights are simply
$\Lambda^\star(1)$, that is
$$\eqalign{
&\{0,\omega_1,\omega_2,\ldots,\omega_{m-1}\},\quad{\rm for}\ a_{m-1},\qquad
\{0,\omega_1,\omega_m\},\quad{\rm for}\ b_m,\cr
&\{0,\omega_1,\omega_2,\ldots,\omega_m\},\quad{\rm for}\ c_m,\qquad
\{0,\omega_1,\omega_{m-1},\omega_m\},\quad{\rm for}\ d_m.\cr}
\efr
For each of the representations $W_a$ (recall from \REP\
that they are in general
reducible for $b$ and $d$ series) one can draw an admissibility diagram
consisting of a set of nodes for each element of $\Lambda^\star(1)$
with $a$ joined to $b$ by an oriented link if $a-b$ is a weight of the
representation. For example, in figures 5 and 6 we give the admissibility
diagrams for the vector spinor and anti-spinor representations of $d_m$ and
for the vector and spinor representations of $c_m$

\sjump
\midinsert
\centerline{
\psfig{figure=ad.eps,height=1.5in}}
\bjump
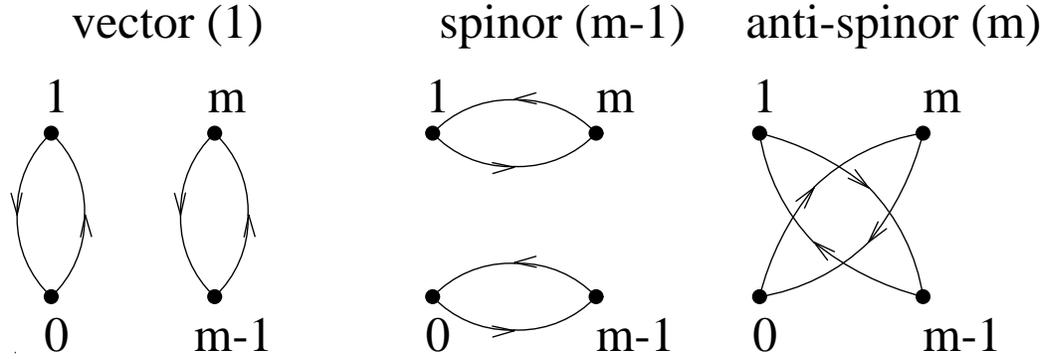
\centerline{5. Admissibility diagrams for $d_m$ with $k=1$.}
\endinsert

\sjump
\midinsert
\centerline{
\psfig{figure=adt.eps,height=1.5in}}
\bjump
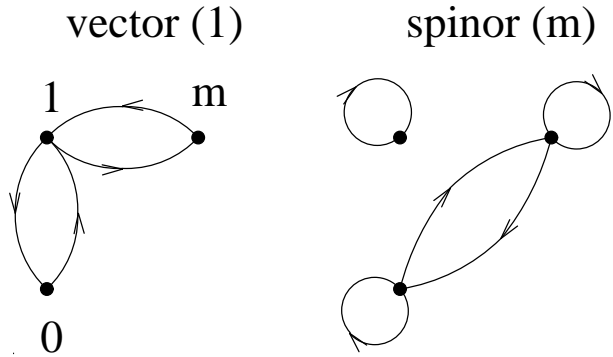
\centerline{6. Admissibility diagrams for $c_m$ and $k=1$.}
\endinsert

One notices immediately a qualitative difference between the admissibility
diagrams of the simply-laced
and non-simply-laced algebras. In the former case the admissibility
diagrams are trivial in the sense that the space of paths on the diagram
from a given starting point and given length is just one-dimensional.
This means that the resulting solution of the YBE is trivial:
$$
{\widetilde S}^{ab}_{(1)}(u)=1.
\efr
This means that the GN-type $S$-matrix for the simply-laced algebras is
related to the PCM-type $S$-matrix:
$$
S^{ab}_{(k)}(u)=S^{ab}_{(k,1)}(u),
\nfr{GNPCM}
and furthermore
$$
S^{ab}_{(1,1)}(u)=S^{ab}_{(1)}(u)=X^{ab}(u),
\nfr{MTB}
the minimal Toda $S$-matrix.

On the contrary, for the non-simply-laced algebras the admissibility
diagrams are not trivial when $k=1$. In other words at $k=1$ the $S$-matrix
is still non-diagonal and a relations like \GNPCM\ and \MTB\ do not hold.

\chapter{Discussion}

We have constructed factorizable $S$-matrices for trigonometric solutions
of the YBE for the vector representations of all the classical Lie algebras.
By appending a suitable CDD factor the $S$-matrices have singularities
corresponding to new states. For the $a_{m-1}$ and $c_m$ algebras, where the
relevant spectral decompositions are known, one can solve the bootstrap
equations to find that the spectrum consists of particles transforming in
the fundamental representations of the algebra. For $b_m$ and $d_m$ the
situation is less clear and in any case the vector particle is not
the fundamental particle; rather one would expect this to be the spinor
(and anti-spinor for $d_m$).

The simplest Ansatz for the $S$-matrix, generalizing that of the Gross-Neveu
model, fails for $c_m$ due to the existence of singularities on the physical
strip which cannot be explained in terms of the spectrum of particles.
A generalized principal chiral model Ansatz, for which the particles
transform in a tensor product of representations, is shown to be consistent
and all singularities on the physical strip can be accounted for.

It has been argued that the $a_{m-1}$ trigonometric $S$-matrices describe
certain integrable deformations of some coset conformal field theories
[\Ref{VF},\Ref{PERT}]. Consider the coset conformal field theory $x_k\times
x_l/x_{k+l}$, for some Lie algebra $x$. There is a particular relevant
operator in the theory of holomorphic dimension
$$
\Delta=1-{g\over k+l+g},
\efr
where $g$ is as before the dual Coxeter number of $x$, which leads to
a massive theory with higher spin integrals of motion. The natural
conjecture is that the $S$-matrix of this integrable theory is precisely
the generalized PCM-type $S$-matrix $S^{ab}_{(k,l)}(u)$ [\Ref{ABL}].
Notice that the $x_1\times x_k/x_{k+1}$ cases are described by the
GN-type $S$-matrix for the simply-laced algebras only (see section 6).
The conjecture could perhaps be placed on a better footing by employing
the technology of the Thermodynamic Bethe Ansatz to investigate the
ultra-violet limit of the theories. The relation of the trigonometric
solutions to the YBE and Bethe Ansatz systems has been considered in
[\Ref{BOOT},\Ref{RW},\Ref{KUN}].

Finally, the $a_{m-1}$ trigonometric $S$-matrices have been proposed to
describe the scattering of solitons in complex $a_{m-1}^{(1)}$ Toda field
theory [\Ref{MSQ},\Ref{MS}]. This has been established via semi-classical
techniques. It is now known that all complex affine Toda theories admit
soliton solutions [\Ref{OUT}], and is natural to ask whether the
trigonometric $S$-matrices for the other algebras describe the scattering
of these solitons. This may be true for the simply-laced algebras, for
which the soliton $S$-matrix would be the generalized GN-type $S$-matrix.
For the non-simply-laced algebras the situation is much less clear.
Experience with the real Toda theories in these cases shows that the
resolution of the problems may be surprisingly subtle and perhaps will
require new types of solution to the YBE appropriate to a set of
particles whose mass ratios depend on a coupling constant.

I would like to thank Patrick Dorey for many conversations on $S$-matrices
and also Gustav Delius for some useful discussions. Also I would like
to thank Niall Mackay for pointing out an error in an earlier version
of the manuscript which allowed for a significant improvement.

\appendix{The spectral decompositions for $a_{m-1}$ and $c_m$}
\def\o{\omega}

In this appendix we derive the spectral decompositions of the $R$-matrices
on the fundamental representations of the $a_{m-1}$ and $c_m$ algebras. We
follow the approach of [\Ref{ZGB}] although our method is also a direct
translation into the quantum group of [\Ref{MK}] which found the spectral
decompositions of the rational $R$-matrices for $c_m$.

The first point to make is an $R$-matrix cannot be associated with any
two representations of the quantum group. It is necessary that the
representations are {\it affinizable\/} in the language of [\Ref{ZGB}].
It can be shown that all the fundamental representations of
$a_{m-1}$ and $c_m$ have this property. However, it is not true for the
fundamental representations of
the other algebras, where the affinizable representations are reducible
in general [\Ref{DRIN},\Ref{OW}].

Now we apply the technology of [\Ref{ZGB},\Ref{MK}]
to find the spectral decompositions
of the $R$-matrices on the fundamental representations. One first
constructs the Tensor Product Graph (TPG). This is a graph is constructed by
letting the irreducible components of the tensor product $V_\mu\ot V_\nu$
be the nodes joined by a link if $V_\lambda$ and $V_\sigma$ have opposite
parity and $V_\sigma\subset{\rm adjoint}\ot V_\lambda$. The parity of an
irreducible component $V_\lambda$ is defined to be $\pm1$ according to whether
$V_\lambda$ appears symmetrically or anti-symmetrically in the tensor product
(in the limit $q\rightarrow1$).

For $V_{\o_a}\otimes V_{\o_b}$ (with $b\geq a$ without loss of generality) of
$a_{m-1}$, using the notation $j,k\equiv V_{\o_j+\o_k}$ and $j\equiv
V_{\o_j}$ we have the TPG
$$
\matrix{a,b&\leftrightarrow&a-1,b+1&\cdots
\leftrightarrow&a-{\rm min}(m-b,a),b+{\rm min}(m-b,a)\cr}.
\efr
For $c_m$ we have for $a+b\leq m$
$$
\matrix{a,b&\leftrightarrow&a-1,b+1&\cdots\leftrightarrow&1,a+b-1&
\leftrightarrow&a+b\cr
\updownarrow&&\updownarrow&&\updownarrow&&\cr
a-1,b-1&\leftrightarrow&a-2,b&\cdots\leftrightarrow&a+b-2&&\cr
\vdots&&\vdots&&&&\cr
\updownarrow&&\updownarrow&&&&\cr
1,b-a+1&\leftrightarrow&b-a+2&&&&\cr
\updownarrow&&&&&&\cr
b-a&&&&&&\cr}
\efr
whilst if $a+b>m$ then the graph truncates at the $(m-b+1)$th column.

The spectral decomposition of the $R$-matrix has the form
$$
\rc^{ab}(x)=\sum_{\mu}\rho_\mu(x){\Bbb P}_\mu,
\efr
where the sum is over the representations
that appear in the tensor product
$V_{\o_a}\otimes V_{\o_b}$ and hence is a sum over nodes of the TPG.
$x$ is the multiplicative spectral parameter. If there is an arrow from
$\nu$ to $\mu$ on the TPG then the coefficients $\rho_\mu(x)$ and
$\rho_\nu(x)$ satisfy
$$
{\rho_\mu(x)\over\rho_\nu(x)}={xq^{I(\mu)/2}-x^{-1}q^{I(\nu)/2}\over
x^{-1}q^{I(\mu)/2}-xq^{I(\nu)/2}},
\nfr{STR}
where $q$ is the deformation parameter of the quantum group and
$I(\mu)=(\mu+2\rho)\cdot\mu$ (where $\rho$ is the sum of the fundamental
weights) is the eigenvalue of the quadratic Casimir on the representation
with highest-weight $\mu$.

The spectral decompositions follow from applying the rule \STR\ recursively
from $\o_a+\o_b$ (for the case of $c_m$ it is important to notice that the
result is independent of the path). With
$$
x=\exp(i\lambda u),\qquad q=-\exp(-i\o),
\efr
one finds \ARSD\ and \OTRSD\ up to an overall multiplicative factor.

\appendix{Zeros and the bootstrap}
\def\ub{\overline u}
\def\S{\widetilde S}

In this appendix we explain how the solution of the bootstrap equations
leads to the zeros in \ARSD\ and \OTRSD.
{}From the spectral decompositions of the $R$-matrices we know that
$$
\S^{ab}(u_{ab}^c)\propto{\Bbb P}_{\o_c},
\efr
for some appropriate value of $u_{ab}^c$,
in which case we say that there is a fusion $ab\rightarrow c$. This also
implies that there are fusions $a\bar c\rightarrow\bar b$ and $b\bar
c\rightarrow\bar a$ where $\bar a$ is the charge conjugate particle (for
$a_{m-1}$ $\bar a=m-a$ whilst for the other algebras $\bar a=a$). The
identity
$$
u_{ab}^c+u_{a\bar c}^{\bar b}+u_{b\bar c}^{\bar a}=2,
\nfr{PFR}
holds between the fusing parameters. The bootstrap equations give the
$S$-matrix elements of $c$ in terms of $a$ and $b$ which
follows from the fusion relation between the $R$-matrices:
$$
\rc^{dc}(u)=\left(I^b\ot\rc^{da}(u+\ub_{a\bar c}^
{\bar b})\right)\left(\rc^{db}(u-\ub_{b\bar c}^{\bar a})\ot
I^a\right),
\nfr{BOOT}
restricted on the left and right to the subspace $V_{\o_c}\subset V_{\o_b}\ot
V_{\o_a}$.
In the above $\ub=1-u$ and $I^a$ is the identity on $V_{\o_a}$. Consider
the case $d=a$ evaluated at $u=-\ub_{a\bar c}^{\bar b}$ then
$$
\rc^{ac}(-\ub_{a\bar c}^{\bar b})=\left(I^b\ot\rc^{aa}(0)\right)
\left(\rc^{ab}(-u_{ab}^c)\ot I^a\right)
\nfr{KLN}
using \PFR. But from the spectral decompositions \ARSD\ and \OTRSD\ one
finds $\rc^{aa}(0)\propto I^a\ot I^a$
and that $\rc^{ab}(-u_{ab}^c)$ has no component
proportional to ${\Bbb P}_{\o_c}$, hence
$$
\rc^{ac}(-\ub_{a\bar c}^{\bar b})=0.
\nfr{ZPO}
We now apply \ZPO\ and \BOOT\ recursively to find zeros of $\rc^{ab}(u)$.
Starting from $\rc^{11}(u)$ (\APR\ and \OTHP) \ZPO\ implies that
$\rc^{12}(-1/tg)=0$. For all the algebras except $a_{m-1}$ crossing
symmetry would imply $\rc^{12}(1+1/tg)=0$ in addition. By the recursive
use of \BOOT\ we find that $\rc^{ab}(u)$ for $b\geq a$ has a set of zeros at
$$
u={1\over tg}(a+b-2j-2k),\quad j=1,2,\ldots,a,\quad k=1,2,\ldots,b-1,
\efr
and for all the algebras except $a_{m-1}$ their crossed values as well.

The remaining zeros in \ARSD\ and \OTRSD\ are accounted for in a different way.
When $a+b>m$ then we saw in appendix A that the TPG gets truncated
which implies an additional set of zeros at
$$
u={1\over tg}(2k+b-a),\quad k=m-b+1,m-b+2,\ldots,a.
\efr

\references

\beginref
\Rref{TOD}{A.E. Arinstein, V.A. Fateev and A.B. Zamolodchikov,
Phys. Lett. {\bf B87} (1979) 389\newline
H.W. Braden, E. Corrigan, P.E. Dorey and R. Sasaki, Nucl. Phys. {\bf B338}
(1990) 689\newline P. Christe and G. Mussardo, Int. J. Mod. Phys. {\bf A5}
(1990) 4581}
\Rref{ABL}{C. Ahn, D. Bernard and A. LeClair,
Nucl. Phys. {\bf B346} (1990) 409}
\Rref{PERT}{A. LeClair, Phys. Lett. {\bf B230} (1989) 103\newline
D. Bernard and A. LeClair, Phys. Lett. {\bf B247} (1990) 309; Nucl. Phys.
{\bf B340} (1990) 721\newline
F.A. Smirnov, Nucl. Phys. {\bf B337} (1990) 156;
Int. J. Mod. Phys. {\bf A4} (1989) 4213\newline
N. Yu Reshetikhin and F. Smirnov, Commun. Math. Phys.
{\bf131} (1990) 157}
\Rref{ZZ}{A.B. Zamolodchikov and Al. B. Zamolodchikov, Ann. Phys. {\bf120}
(1979) 253}
\Rref{JIM}{M. Jimbo, Lett. Math. Phys. {\bf10} (1985) 63;
Lett. Math. Phys. {\bf11} (1986) 247; Int. J. Mod. Phys. {\bf A4} (1989) 3759}
\Rref{DRIN}{V.G. Drinfel'd, Sov. Math. Dokl. {\bf32} (1985) 254}
\Rref{MS}{T.J. Hollowood, Nucl. Phys. {\bf B384} (1992) 523}
\Rref{MSQ}{T.J. Hollowood, Int. J. Mod. Phys. {\bf A8} (1993) 947;
`{\sl A quantum group approach to
constructing factorizable $S$-matrices\/}', Oxford University preprint
OUTP-90-15P}
\Rref{BOOT}{E. Ogievetsky, P. Wiegmann and N. Reshetikhin, Nucl.
Phys. {\bf B280} (1987) 45}
\Rref{NLS}{G.W. Delius, M.T. Grisaru and D. Zanon, Nucl. Phys. {\bf B382}
(1992) 365}
\Rref{VF}{H.J. de Vega and V.A. Fateev, Int. J. Mod. Phys. {\bf A6} (1991)
3221}
\Rref{JMO}{M. Jimbo, T. Miwa and M. Okado, Comm. Math. Phys. {\bf116} (1988)
507}
\Rref{WDA}{M. Wadati, T. Deguchi and Y. Akutsu, Phys. Rep. {\bf180} (1989)
247}
\Rref{MK}{N.J. MacKay, J. Phys. A: Math. Gen. {\bf25} (1992) L1343}
\Rref{GN}{B. Berg, M. Karowski, P. Weisz and V. Kurak, Nucl. Phys. {\bf B134}
(1978) 125}
\Rref{OUT}{D.I. Olive, N. Turok and J.W.R. Underwood, `{\sl Solitons and
the energy-momentum tensor for affine Toda theory}', preprint
Imperial/TP/91-92/35 Swansea SWAT/3}
\Rref{ZGB}{R.B. Zhang, M.D. Gould and A.J. Bracken, Nucl. Phys. {\bf B354}
(1991) 625}
\Rref{OW}{E. Ogievetsky and P.B. Wiegmann, Phys. Lett. {\bf168B} (1986) 360}
\Rref{RW}{N. Yu. Reshetikhin and P.B. Wiegmann, Phys. Lett.
{\bf189B} (1987) 125}
\Rref{GEP}{D. Gepner, `{\sl On RSOS models associated to Lie algebras and
RCFT}', Caltech preprint}
\Rref{OSD}{S. Coleman and H.J. Thun, Commun. Math. Phys. {\bf61} (1978) 31
\newline H.W. Braden, E. Corrigan, P.E. Dorey and R. Sasaki, Nucl. Phys.
{\bf B356} (1991) 469}
\Rref{KUN}{A. Kuniba, Nucl. Phys. {\bf B389} (1993) 209}
\Rref{CDS}{E. Corrigan, P.E. Dorey and R. Sasaki, `{\sl On a generalized
bootstrap principle\/}', Durham preprint DTP-93/19}
\endref
\ciao